\documentclass[11pt,preprint]{aastex}
\usepackage{natbib}

\def\t0{\theta_{\circ}}

\def\be{\begin{equation}}
\def\en{\end{equation}}

\def\msun{M_{\sun}}
\def\rsun{R_{\sun}}
\def\lsun{L_{\sun}}
\def\msunyr{{\rm{M_{\sun} \; yr^{-1}}}}

\def\mdot{\dot{M}}

\def\curf{{\cal F}}

\def\h2{H$_2$}
\def\rmxaa{Rev. Mexicana Astron. Astrofis.}

\begin{document}

\title
{Accretion Rates for T Tauri Stars Using Nearly Simultaneous  Ultraviolet and Optical Spectra}

\author{Laura Ingleby\altaffilmark{1}, Nuria Calvet\altaffilmark{1}, Gregory Herczeg\altaffilmark{2}, Alex Blaty\altaffilmark{1}, Frederick Walter\altaffilmark{3}, David Ardila\altaffilmark{4}, Richard Alexander\altaffilmark{5}, Suzan Edwards\altaffilmark{6}, Catherine Espaillat\altaffilmark{7}, Scott G. Gregory\altaffilmark{8,}\altaffilmark{9}, Lynne Hillenbrand\altaffilmark{8}, Alexander Brown\altaffilmark{10}
}

\altaffiltext{1}{Department of Astronomy, University of Michigan, 830 Dennison Building, 500 Church Street, Ann Arbor, MI 48109, USA; lingleby@umich.edu, ncalvet@umich.edu, ablaty@umich.edu}
\altaffiltext{2}{The Kavli Institute for Astronomy and Astrophysics, Peking University, Yi He Yuan Lu 5, Hai Dian Qu, 100871 Beijing, P. R. China}
\altaffiltext{3}{Stony Brook University, Stony Brook NY 11794-3800, USA}
\altaffiltext{4}{NASA Herschel Science Center, California Institute of Technology, Mail Code 100-22, Pasadena, CA 91125, USA}
\altaffiltext{5}{Department of Physics \& Astronomy, University of Leicester, University Road, Leicester, LE1 7RH, UK}
\altaffiltext{6}{Department of Astronomy, Smith College, Northampton, MA 01063, USA}
\altaffiltext{7}{NASA Sagan Postdoctoral Fellow.  Harvard-Smithsonian Center for Astrophysics, 60 Garden Street, MS-78, Cambridge, MA 02138, USA}
\altaffiltext{8}{California Institute of Technology, Department of Astrophysics, MC 249-17, Pasadena, CA 91125, USA}
\altaffiltext{9}{School of Physics \& Astronomy, University of St. Andrews, St. Andrews, KY16 9SS, UK}
\altaffiltext{10}{Center for Astrophysics and Space Astronomy, University of Colorado, Boulder, CO 80309-0389, USA}



\begin{abstract}
We analyze the accretion properties of 21 low mass T Tauri stars using a dataset of contemporaneous near ultraviolet (NUV) through optical observations obtained with the \emph{Hubble Space Telescope} Imaging Spectrograph (STIS) and the ground based Small and Medium Aperture Research Telescope System (SMARTS), a unique dataset because of the nearly simultaneous broad wavelength coverage.  Our dataset includes accreting T Tauri stars (CTTS) in Taurus, Chamaeleon I, $\eta$ Chamaeleon and the TW Hydra Association.  For each source we calculate the accretion rate ($\mdot$) by fitting the NUV and optical excesses above the photosphere, produced in the accretion shock, introducing multiple accretion components characterized by a range in energy flux (or density) for the first time.  This treatment is motivated by models of the magnetospheric geometry and accretion footprints, which predict that high density, low filling factor accretion spots co-exist with low density, high filling factor spots.  By fitting the UV and optical spectra with multiple accretion components, we can explain excesses which have been observed in the near infrared.  Comparing our estimates of $\mdot$ to previous estimates, we find some discrepancies; however, they may be accounted for when considering assumptions for the amount of extinction and variability in optical spectra.  Therefore, we confirm many previous estimates of the accretion rate.   Finally, we measure emission line luminosities from the same spectra used for the $\mdot$ estimates, to produce correlations between accretion indicators (H$\beta$, Ca II K, C II] and Mg II)  and accretion properties obtained simultaneously.

\end{abstract}


\section{ Introduction}
\label{intro}
 Classical T Tauri stars (CTTS) are pre-main sequence objects that are accreting gas from their circumstellar disks.  
The currently accepted paradigm is magnetospheric accretion, where the circumstellar disk is truncated at a few stellar radii by the stellar magnetosphere \citep{hartmann94,bouvier95,johnskrull99,johnskrull02,muzerolle98,muzerolle01},  with strong magnetic field strengths of a few kG \citep{johnskrull00}.  The gas at the truncation radius is channeled along the magnetic field lines at nearly free fall velocities, until it impacts the stationary photosphere, producing an accretion shock.  Accreting sources are identified by broad emission lines (like H$\alpha$, Ca II and FUV lines of C IV and He II), tracing the fast moving material in the accretion streams \citep{edwards94,muzerolle98,ardila12}, or by an ultraviolet (UV) and $U$ band excess over the stellar photosphere, produced by hot gas in the accretion shock \citep[hereafter CG98]{calvet98}.

CTTS are known to vary in brightness over short timescales, due to changes in the accretion luminosity and the modulation of spots on the stellar surface \citep{herbst94,alencar12}.  In addition, a spread in the accretion rate is observed for sources of the same age \citep{hartmann98,calvet05,sicilia10}.  Therefore, the accretion properties of young stars are best characterized by studying a large sample, including non-accreting T Tauri stars (WTTS) for comparison.  WTTS are no longer accreting, but they retain strong magnetic fields, due to magnetic dynamo effects, which heat the chromosphere.  The ramped up activity in the chromosphere is seen as a UV excess over a dwarf star photosphere \citep{houdebine96} and emission in lines of  hydrogen and calcium, which are the same diagnostics of magnetospheric accretion; however, the chromospheric contribution is much weaker \citep{ingleby11b}.  Due to the similarity in tracers, it is essential to use a WTTS template when estimating the accretion rate onto the star ($\mdot$), especially for low $\mdot$ objects where the accretion emission is comparable to that from the active chromosphere.  Until now, high signal to noise UV spectra of WTTS were not available so dwarf spectra, with lower chromospheric activity levels, took their place as stellar templates, resulting in overestimated UV excesses.  Accurate $\mdot$s are vital for the understanding of disk physics.  The accretion rate provides information about the surface density of the inner circumstellar disk; for the lowest accretors, the accretion properties reveal the characteristics of the final stages of the inner disk, shortly before it is dissipated  \citep{dalessio99,ingleby11b}.

Initial attempts to fit the UV excess produced in the accretion shock assumed accretion spots on the stellar surface were characterized by either a single temperature and density in treatments of the shock as a slab \citep{valenti93,gullbring98,herczeg08,rigliaco12}, or a single energy flux in the accretion column for accretion shock models \citep[see Section \ref{subshock}]{calvet98}.  These models indicated that the accretion spots cover a small fraction of the stellar surface, tenths to a few percent.  Recent models of the magnetosphere, calculated to reproduce spectro-polarimetric observations, revealed that the magnetic field geometry of accreting stars is complex, including high order and tilted fields \citep{donati08,gregory11,gregory12}.  The accretion footprints on the stellar surface which result from field lines of varying strengths and geometries are themselves complex, ranging in size and density of the shocked material, with filling factors which may exceed 10\% of the stellar surface \citep{long11}.  These large filling factors have yet to be reproduced when fitting the accretion excess without significantly over-estimating the UV fluxes.

In addition to the large filling factors, near infrared (IR) veiling observations are not explained by current accretion shock models.  Veiling occurs when the excess emission produced in the shock fills in, or ``veils", photospheric absorption lines, causing them to appear shallow when compared to a WTTS template \citep{hartigan91,valenti93,johnskrull99,dodin12}.  Veiling at progressively redder wavelengths can be found in the literature, out to 1 $\mu$m \citep{basri90,hartigan91,white04,edwards06}.  More recent results on near IR veiling were discussed in \citet{fischer11} and \citet{mcclure12}, who showed that the amount of veiling becomes nearly constant from 0.8 to 1.4 $\mu$m.    For an accretion shock spectrum which peaks in the UV and decreases toward long wavelengths, veiling should be negligible in the near-IR.  The spectrum of the disk, assuming it is produced by circumstellar dust at the dust sublimation temperature around 1400 K, also has little contribution to the veiling near 1 $\mu$m and does not begin to contribute significantly until $\sim$2 $\mu$m \citep{fischer11}.    Long wavelength veiling may originate in a cool accretion component, cooler than that which describes the UV excess \citep{calvet98,white04}; however, \citet{fischer11} found that in some cases
the surface area of the cool accretion
column required to explain the veiling was larger than
the stellar surface, so they suggested that
the veiling came from hot gas in the inner circumstellar disk.
On the other hand, by varying the sublimation temperature
of the dust over a range consistent with different materials,
\citet{mcclure12} showed that the veiling between
0.8 and 2.32 microns could be explained by the combined
emission of dust at the sublimation radius
and a cool accretion column with a
reasonable filling factor.

In this paper, we assume that 1 $\mu$m veiling is produced by cool accretion components and use the accretion shock models described in CG98 to fit UV and optical observations of a large sample of CTTS.  The sample is a part of the large  \emph{Hubble Space Telescope} (\emph{HST}) program, Disks, Accretion and Outflows of T Tau stars (DAO; PI G. Herczeg) which compiled a dataset for each source covering a long range in wavelength with observations as close to simultaneous as possible.  Observations included \emph{HST} far UV (FUV), near UV (NUV) and optical spectra obtained with the Cosmic Origins Spectrograph (COS) and the Space Telescope Imaging Spectrograph (STIS).   We focus on the NUV and optical observations which are ideal for measuring accretion excesses, whereas the FUV spectrum is complicated by molecular emission, both in lines and the continuum \citep{bergin04,ingleby11b,france11}.  Attempts were made to contemporaneously (within a few nights of the \emph{HST} observations) obtain additional optical spectra in order to observe H$\alpha$, which was not covered in the STIS spectra; these attempts were successful for over half of the sample (Section \ref{ground}).

Here, we improve upon treatments of the accretion shock as a single spot on the stellar surface by including multiple accretion components, a scenario more physically accurate given the geometry of the magnetosphere.   With the added components, we attempt to explain long wavelength veiling by including cooler accretion columns, as suggested by \citet{white04} and CG98, from which the emission peaks toward redder wavelengths.  In Section \ref{obs} we discuss the data used in this paper, both new observations and some previously published.  Section \ref{calc} describes the process we use for calculating accretion rates and our results are presented in Section \ref{results}.  Finally, in Section \ref{hidden} we explore the range of accretion rates possible, assuming that some accretion luminosity is undetectable above the star and in Section \ref{correlations} we show how our new results affect correlations between accretion indicators and $\mdot$, which are commonly used when UV observations are not feasible.

\section{Sample and Observations}
\label{obs}

\subsection{Description of Sample}
The sample consists of 21 low mass CTTS and 4 WTTS, primarily from the large \emph{HST} program GO 11616 (PI: G. Herczeg), including three CTTS observed earlier with STIS, BP Tau and TW Hya (GO program 9081; PI. Calvet) and LkCa 15 (GO program 9374; PI. Bergin).  The total DAO sample is larger and will be presented in \citet{herczeg13}, but here we select those sources which are low mass (spectral types late G to early M).  The high mass sources in the sample lack accurate templates which are necessary for our analysis.  The sources included in this paper are in the 1--2 Myr old Taurus Molecular Cloud \citep{kenyon95}, the 2--3 Myr old Chamaeleon I star forming region \citep{gauvin92}, the $\sim$9 Myr $\eta$ Chamaeleon region \citep{lawson04}, the 10 Myr TW Hydra Association \citep{webb99}, and the 16 Myr Lower Centaurus Crux subgroup \citep{pecaut12}.  Each of these regions is characterized as having isolated low mass star formation, with relatively low extinction, so the sources are not affected by remaining molecular cloud material or nearby high mass stars.  The WTTS and CTTS studied in the paper are listed in Tables \ref{tabwtts} and \ref{tabctts}.

The majority of the sources are single stars; however there are a few wide binaries.  DK Tau A, HN Tau A and RW Aur A all have companions at large separations  of $>$1.4\arcsec or $\sim$200 AU \citep{kraus11,white01}, resolved by STIS.  CS Cha is a spectroscopic binary \citep{guenther07,nguyen12}, surrounded by a circumbinary disk \citep{takami03,nagel12}.  Several of the sources have disks which show evidence for gaps or holes in their infrared spectral energy distributions, including CS Cha, DM Tau, GM Aur and LkCa 15, but all of these sources are still accreting \citep{strom89,espaillat07,espaillat07b,espaillat10}.

\subsection{\emph{HST} Observations}
\label{hst}

Observations were obtained with STIS between 2009 and 2012.  STIS NUV observations used the MAMA detector and the G230L
grating providing spectral coverage from 1570 to 3180 {\AA} with
R$\sim$500-1000.  Optical observations were completed during the same
orbit as the NUV using the G430L grating which covers 2900-5700 {\AA}
with R$\sim$530--1040, resulting in almost simultaneous NUV to optical
coverage with STIS.  The bright source, CV Cha, was observed with the echelle grating in the NUV (E230M), which we use in place of the G230L spectrum.  

The low-resolution STIS spectra were calibrated with custom written
IDL routines following the procedures described in the STIS data
handbook.  The wavelengths were calibrated from the location of
identified emission lines within the spectrum, and fluxes were calibrated
from spectra of WD 1337+705 in the NUV and HIP 45880 in the optical.
The flux calibration also includes a wavelength-dependent aperture
correction.  Two of the STIS spectra (those of RECX 1 and RECX 11) were previously discussed in \citet{ingleby11b}

\subsection{Ground Based Observations}
\label{ground}
We also obtained low dispersion spectra of much of the sample, covering the
H$\alpha$ line, using the
Small and Medium Aperture Research Telescope System (SMARTS) 1.5m
telescope at the Cerro Tololo Inter-American Observatory (CTIO) with spectral coverage between $\sim$5600 -- 7000 {\AA}.  Sources were observed with SMARTS for a few days before and after the \emph{HST} observations when possible, providing contemporaneous observations of H$\alpha$.  The sample of CTTS with SMARTS coverage includes AA Tau, CS Cha, CV Cha, DE Tau, DM Tau, DN Tau, FM Tau, PDS 66, RECX 11, RECX 15, TWA 3a and V836 Tau.  For the following analysis, we used the epoch of SMARTS spectra which was nearest in time to the \emph{HST} observations.  The reduction of SMARTS spectra was described in \citet{ingleby11b}.  The flux calibration of the SMARTS spectra was not completely accurate due to flux losses in the slit and changes in the observing conditions.  To best estimate the absolute fluxes, we scaled the SMARTS spectra to the \emph{HST} optical spectra in the region where the two overlap, assuming that the shape of the spectrum is accurate.  We assessed the degree of variability in the shape of our SMARTS spectra by computing the standard deviation of the fluxes in the different epochs and dividing it by the median of the observed spectra.  With spectral shape variability on the order of 10--25\% for our sample (whether due to observing conditions or intrinsic variability), there is little error introduced by fitting only one SMARTS optical spectrum per source.  

Due to the low resolution of the SMARTS data, when available, we supplemented the observations with non-simultaneous high resolution spectra obtained with MIKE (Magellan Inamori Kyocera Echelle) on the Magellan-Clay telescope at Las Campanas Observatory in Chile.  MIKE has a spectral coverage of 4800--9000 {\AA} and resolution of R$\sim$35,000. The data were reduced using the Image Reduction and Analysis Facility (IRAF) tasks CCDPROC, APFLATTEN, and DOECSLIT \citep{tody93}. MIKE spectra were used to calculate V band veiling, which would not be accurate at the low resolution of the SMARTS data (Section \ref{sectveil}).  We use MIKE spectra to obtain $V$ band veiling for CS Cha, DM Tau, LkCa 15, RECX 11, RECX 15 and TWA 3a. With its location in Chile, Taurus observations are difficult and therefore much of our sample was not covered with MIKE.  Two sources were observed with the echelle spectrograph on the SMARTS 1.5m telescope, CV Cha and PDS 66, allowing for veiling estimates.

Additional observations were obtained with CRIRES, the CRyogenic high-resolution InfraRed Echelle Spectrograph on the Very Large Telescope (VLT).  CRIRES has a spectral resolution up to $10^5$ in the range of 1 to 5 $\mu$m.  These spectra are used in this paper to calculate veiling for a small number of sources in our sample with unknown long wavelength veiling measurements (see Section \ref{sectveil}).  A description of the CRIRES data reduction may be found in \citet{ingleby11b}.   A log of all HST and ground based observations is given in Table \ref{tabobs}.

\section{Calculating Accretion Rates}
\label{calc}

\subsection{Stellar Template: The Active Chromosphere}
Enhanced chromospheric activity is characteristic of stars which have not yet reached the main sequence \citep{bertout89,guinan03}.  A higher level of activity is inferred from lines like H$\alpha$ and the Ca II infrared triplet, which are in emission and also variable in young stars \citep{galvez09}.  WTTS also show excess emission at UV through blue optical wavelengths compared to dwarf stars \citep{houdebine96}.  Since H$\alpha$, Ca II emission lines and the UV excess are also typical tracers of accretion, emission from the active chromosphere must be taken into account when the accretion luminosity is estimated from any of the aforementioned indicators.  While chromospheric emission is not a large contaminant for strong accretors, it is extremely significant when $\mdot$ is low.  In \citet{ingleby11b} we demonstrated how the emission from an active chromosphere can mask all evidence of an accretion shock excess for the lowest accretors, though line profiles observed at high resolution revealed complex absorption features produced by the accretion flows.  

\begin{figure}[htp]
\plotone{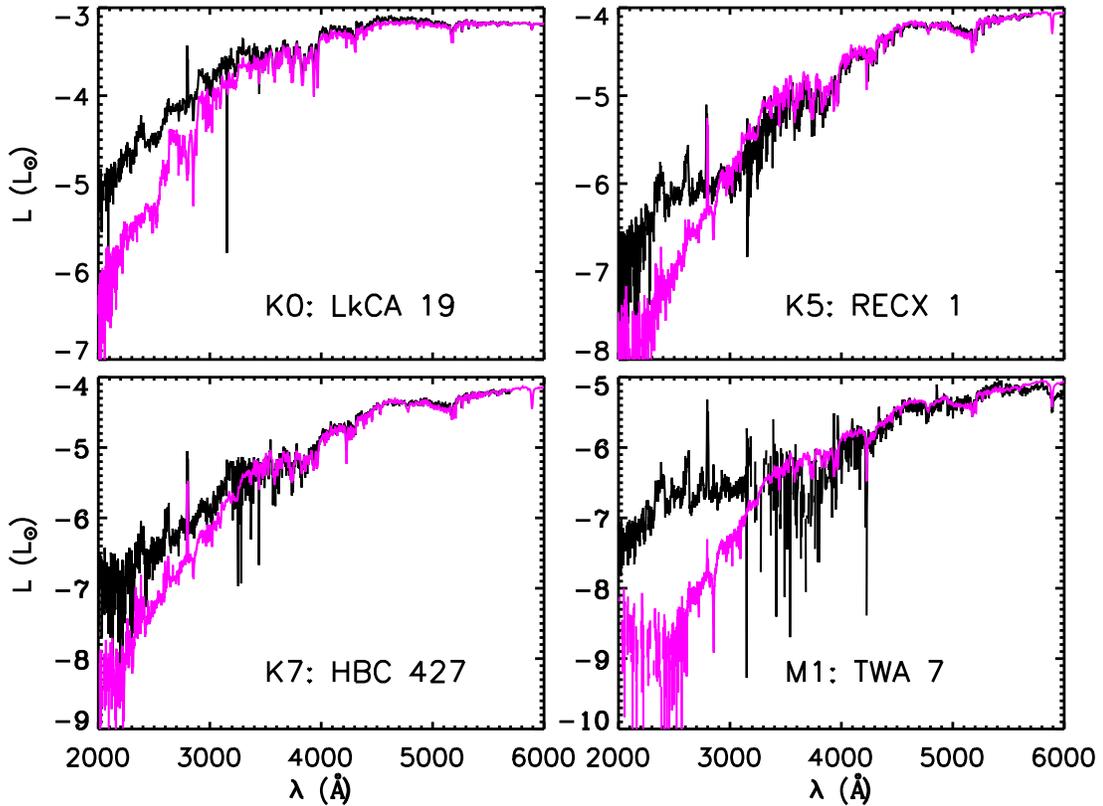}
\caption[Comparison of WTTS and dwarf stars]{Comparison of WTTS and dwarf stars.  In each panel the black line is the WTTS observed as part of the DAO sample and the magenta line is a dwarf standard of the same spectral type, taken from the STIS Next Generation Spectral Library.  The WTTS excess in the NUV is produced by an active chromosphere in young stars.}
\label{dwarf}
\end{figure}

Previous estimates of $\mdot$ have mainly relied on dwarf photospheres as templates against which to measure the U band or UV excess \citep{romaniello04}.  A better estimate of the excess  could be found using a WTTS template \citep{valenti93} but, until now, WTTS with good signal in the UV continuum were not available.  As part of the DAO sample, WTTS were included to act as templates, covering the range of spectral types of the accreting sample (Table \ref{tabwtts}).  Figure \ref{dwarf} shows the WTTS templates compared to main sequence dwarf stars with the same spectral type; dwarf spectra were scaled to the WTTS at 5500 {\AA}.  STIS observations of main sequence stars were taken from the \emph{HST} Next Generation Spectral Library \citep{heap07}.   We compared the luminosity of the WTTS to that of the dwarf between 2000 and 3000 {\AA} and found that the WTTS had NUV luminosities $\sim$3$\times$ higher than the dwarf stars.  \citet{findeisen11} showed that WTTS of a given age and spectral type exhibit a range in UV fluxes, so the example WTTS shown in Figure \ref{dwarf} may not exactly represent the chromosphere for each individual source, but they are the best templates to date.   In the following analysis, we use the WTTS with the closest spectral type match to each CTTS as the stellar template.

\subsection{Veiling By Shock Emission}
\label{sectveil}
The excess emission produced in the shock can contribute significantly to the total luminosity of the CTTS, making it difficult to determine the luminosity of the star itself.  However, veiled absorption lines provide a diagnostic of the relative contributions from the star and shock \citep{gullbring98}.  Veiling occurs when excess emission is added to the spectrum of the star, filling in the photospheric absorption lines \citep{hartigan89,hartigan91}, so by comparing the depth of absorption lines in the CTTS to a WTTS, we have an estimate of the veiling continuum.  There are some uncertainties in this method; in particular \citet{gahm08} and \citet{dodin12} show that emission lines can also fill in photospheric absorption lines, so treating the veiling emission as a continuum may not be correct; however, this effect will likely only be important for sources with the strongest emission line spectra, or highest veiling \citep{petrov11}.

\begin{figure}[htp]
\plotone{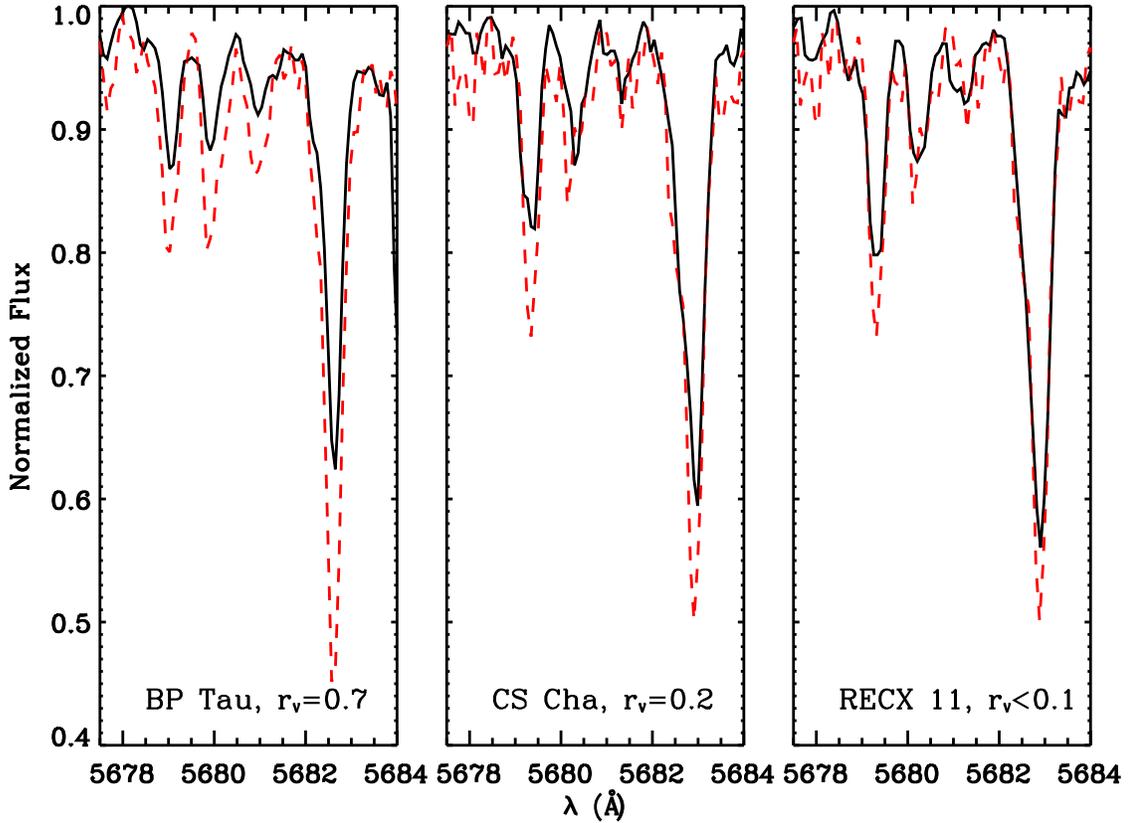}
\caption[Veiling in MIKE spectra] {Veiling in MIKE spectra.  The panels show three CTTS (solid, black) observed with MIKE which have different degrees of veiling compared to a WTTS of the same spectral type (dashed, red).  The CTTS are BP Tau (left) with spectral type of K7, CS Cha (middle) with spectral type K6 and RECX11 (right) with spectral type of K5.  BP Tau has significantly more veiling than RECX 11, observed as shallower absorption lines.  Typical errors on the veiling are $\pm$0.1.}
\label{veil}
\end{figure}

The continuum flux responsible for the veiling of absorption
lines is added to the flux of the photosphere, here taken to be
a WTTS,  to produce the observed spectrum. So, 
veiling at one wavelength allows us to estimate the intrinsic
photospheric flux at that wavelength from the observed
spectrum. If we have an accurate
photospheric template, this scaling provides the stellar
component flux over the entire spectrum. 
When available, we used the veiling at $V$, published in \citet{edwards06}, which provides a compilation from the literature.  For a small sample of the sources with unpublished veiling (CS Cha, CV Cha, DM Tau, LkCa 15, PDS 66, RECX 11, RECX 15 and TWA 3a) we calculated $r_V$ from high resolution optical spectra obtained with MIKE or the SMARTS echelle.   For templates, we used echelle spectra of WTTS with the same spectral type as each of our sources.  We then added a continuum excess to the WTTS spectrum until the depth of the absorption lines matched those in the CTTS, giving us the veiling continuum.  We show three examples of different degrees of veiling observed in our MIKE spectra in Figure \ref{veil}.  
  
Based on the veiling measurements at $V$ band, we scaled our WTTS templates to each CTTS using,
\begin{equation}
F_{V,WTTS}=F_{V,CTTS}/(1+r_{V}),\\
\end{equation}
where $F_{V,WTTS}$ and  $F_{V,CTTS}$ are the continuum fluxes of the WTTS and CTTS at $V$, respectively, and $r_{V}=F_{V,veil}/F_{V,WTTS}$, is the veiling at V, where $F_{V,veil}$ is the excess continuum emission added to the photospheric spectrum at $V$.  It is important to note that none of our $V$ band veilings are simultaneous to the STIS UV and optical spectra.  This introduces uncertainties in our estimates of $\mdot$ because CTTS accretion properties are known to be variable and as $\mdot$ (and the excess continuum emission) change, so does the veiling \citep{alencar12}.  Ideally, $\mdot$s should be measured using simultaneous long wavelength veiling information to accurately assess the flux contribution from the star and measure the accretion excess.

According to accretion shock models, veiling should drop to 0 in the near IR as the shock spectrum is peaked in the UV (CG98); however, non-zero veiling at long wavelengths has been measured in some CTTS, which we attempt to explain in this paper.  Veilings at 1 $\mu$m were found in the literature for a subset of 10 CTTS in this sample \citep{edwards06}.  For another five sources, we measured veiling near 1 $\mu$m using our CRIRES spectra with either the WTTS RECX 1 as a template (for CS Cha, LkCa 15 and RECX 11) or using a Near-Infrared Spectrograph (NIRSPEC) spectrum of a WTTS \citep{edwards06}.  The NIRSPEC template was used to calculate 1 $\mu$m veiling for DM Tau, FM Tau and IP Tau, after convolving the CRIRES observations to the lower resolution of NIRSPEC.  For CV Cha, PDS 66, RECX 15 and TWA3a, we did not have an appropriate template to calculate the veiling, so we used the relation, $r_Y=0.5 \times r_V$ \citep{fischer11} to obtain $r_Y$.  We also used this relation to estimate $r_V$ for IP Tau, which was not in the literature and for which we did we have high resolution optical spectra; however we did have CRIRES spectra to calculate $r_Y$.  All values of $r_Y$ and $r_V$ are listed in Table \ref{tabctts}

\subsection{Extinction and Stellar Parameters}
A large source of error in our estimates of $\mdot$ comes from the assumed amount of extinction, or $A_V$.  Our STIS dataset did not provide the resolution necessary to measure veiling in the optical and determine $A_V$ based on the veiling, as in \citet{gullbring98}, so we use extinction estimates from the literature.  Table \ref{tabav} shows that $A_V$ values for a given source vary widely in the literature.  Some of the discrepancies in $A_V$ may be due to true variability in the extinction, perhaps by inhomogeneities in the intervening molecular cloud or warps in the circumstellar disk.  Variable accretion hot spots can also affect the colors of the star, affecting the $A_V$ determinations \citep{carpenter01,carpenter02}.  \citet{gullbring98} showed that calculating $A_V$ by comparing the colors of reddened stars to standard photospheric colors results in different values depending on the colors used.  In particular, they found that $V-I$ colors are sensitive to the spectral type classification.

A recently developed method to estimate $A_V$ uses IR veiling estimates to determine the shape of the reddened photosphere below the excess and correct accordingly; however, this method has only been applied to a few sources in our sample (see Fischer et al. 2011 and McClure et al. 2013).   Both Fischer et al. (2011) and McClure et al. (2012) note that their estimates are higher than the $A_V$s from frequently quoted sources, like \citet{kenyon95}.  There are still discrepancies among estimates of $A_V$ for the same source, using the IR veiling method (Table \ref{tabav}).     For consistency, we use the $A_V$ estimates from \citet{furlan09,furlan11} when correcting Taurus and Chameleon I sources for extinction.  \citet{furlan09,furlan11}, computed extinction values by comparing observed $V-I$, $I-J$, or $J-H$ colors to expected photospheric colors in \citet{kenyon95}, using the reddest colors available to minimize the impact from the shock excess which is typically strongest in the blue.  For the remaining sources (TW Hya, TWA 3a, RECX 11 and RECX 15) we assume $A_V$=0 \citep{webb99,luhman04b}.  Errors of $\pm0$.5 in $A_V$ can lead to an uncertainty up to 1 order of magnitude in $\mdot$ making extinction estimates the largest source of error in our calculations.

The STIS and SMARTS spectra were de-reddened using the reddening law of \citet{whittet04}.  UV emission is extremely sensitive to reddening assumptions, with $A_{\lambda}>2\times A_V$ near 2500 {\AA} and corrections are complicated by uncertain UV reddening laws.  \citet{calvet04} compared UV reddening laws and found that of \citet{whittet04} was appropriate for sources in environments like the Taurus Molecular Cloud.  We calculated  the stellar luminosity from the flux in the $J$ band of the photosphere using the bolometric correction of \citet{kenyon95}. The photospheric $J$ band flux was obtained by scaling de-reddened 2MASS $J$ magnitudes by the veiling at  1 $\mu$m.  Although there is an excess above the photosphere at $J$, with known veiling we separate the $J$ flux of the star from that of any continuum excess and use the flux from the star alone when calculating the luminosity.   We then used the \citet{siess00} evolutionary tracks to determine the masses of the sources in the sample and estimate radii from the luminosities.  The stellar properties for our WTTS and CTTS are listed in Tables \ref{tabwtts} and \ref{tabctts}, respectively.

\subsection{Accretion Shock Model}
\label{subshock}
A full description of the model used to characterize the accretion shock
can be found in CG98 but here we review the main points.  The
current picture of material accretion in the inner disk of CTTS is
magnetospheric accretion.  Columns of accreting material fall onto the star along the
magnetic field lines traveling at 
the free fall velocity, $v_s$,
hit the stationary
photosphere and create a shock.  The velocity of the material is given
by 
\be
\label{vs}
v_s=\left(\frac{2GM_{\ast}}{R_{\ast}}\right)^{1/2}\left(1-\frac{R_{\ast}}{R_i}\right)^{1/2}\\,
\en 
where $M_{\ast}$ and $R_{\ast}$ are the stellar mass and radius, respectively, and $R_i$ (assumed to be 5$\;R_{\ast}$) is the radius at which
the magnetosphere truncates the disk (CG98).

The model is simplified by assuming that
the accretion column has a plane parallel geometry and is perpendicular
to the stellar photosphere.  The shock formed at the base of the
accretion column reduces the velocity of the infalling material in
order for it to join the star at the photosphere, converting the
kinetic energy into thermal energy, and causing the temperature 
to increase sharply.  In the strong shock approximation, which
is used because the material is traveling at high velocities, the
temperature immediately after the shock is described by 
\be
T_s=8.6\times10^5\,K\,\left(\frac{M_{\ast}}{0.5\msun}\right)\left(\frac{R_{\ast}}{2\rsun}\right)^{-1}
\en 
(CG98).  At these high temperatures, the shocked material emits
soft X-rays into the pre-shock, post-shock and the
photosphere below the shock. This radiation heats the
material in these regions causing it to emit the observed excess
continuum emission.

\begin{figure}[htp]
\plotone{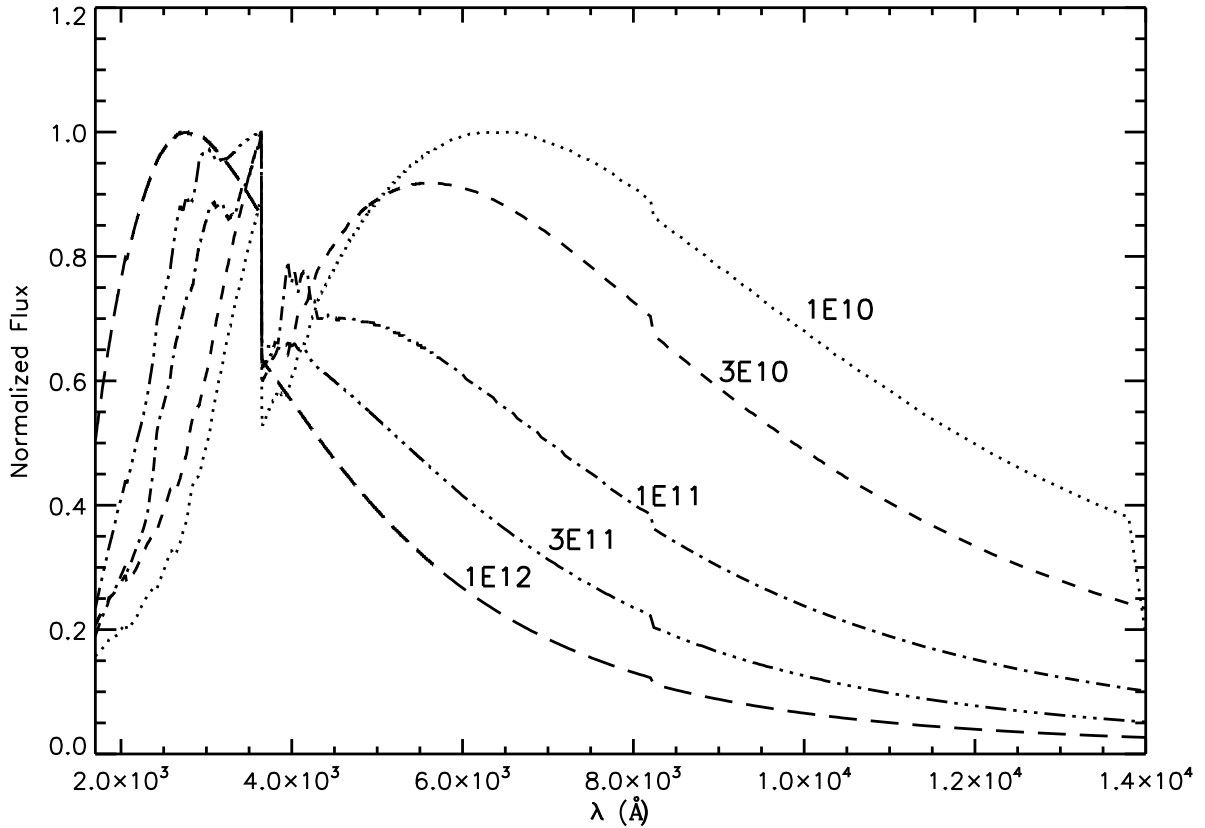}
\caption[Fluxes of accretion shock models with varying energy flux ($\curf$)]{Fluxes of accretion shock models with varying energy flux ($\curf$).  Each line shows the emission from an accretion column with a different value of $\curf$, normalized to unity.  The energy flux of the column is listed next to each spectrum, spanning the range from $\curf=10^{10}-10^{12}\;\rm{ erg\; s^{-1}\; cm^{-2}}$.  The peak of the emission shifts to longer wavelengths as $\curf$ decreases.  }
\label{shocks}
\end{figure}

In this treatment, the emission from the accretion column is characterized by two parameters,
$\curf$ and $f$, which are the total energy flux in the accretion
column ($\curf=1/2\rho v_s^3$, where $\rho$ is the density of the material in the accretion column) and the filling factor (or fraction of the surface of the
star which is covered by the accretion hot spot), respectively.  
Changes in the filling factor, $f$, cause the shock spectrum to increase or decrease independent of wavelength, in essence scaling the luminosity of the emission.  $\curf$ acts to change the wavelength of the peak of the accretion shock emission.  Figure \ref{shocks} shows how the shock spectrum, normalized by the maximum flux of each spectrum, changes for different values of $\curf$, assuming typical values for $M_{\ast}$ and $R_{\ast}$ of 0.5 $\msun$ and 2.0 $\rsun$, respectively.  As $\curf$ increases, more energy is deposited on the stellar surface, increasing the temperature of the photosphere below the shock. This region behaves as the photosphere of a star with an earlier spectral type than that of the undisturbed photosphere.  High $\curf$ columns have temperatures up to 9000 K, whereas low $\curf$ columns are cooler, around 5000--6000 K, producing the wavelength shift between columns with different $\curf$s.  For the lowest value of $\curf$, the emission peaks around 6000--7000 {\AA} whereas for the model with the highest $\curf$, the emission peaks between 2000 and 3000 {\AA}.  

 We use this property of the accretion column emission to explain excesses at both short and long wavelengths when attempting to fit the DAO observations.  In the following text we refer to accretion columns with energy fluxes of $\curf\le10^{11}\;\rm{erg\; s^{-1}\; cm^{-2}}$ as ``low $\curf$" columns and those characterized by  $\curf>10^{11}\;\rm{erg\; s^{-1}\; cm^{-2}}$ as ``high $\curf$" columns.  $\curf$ depends on the density and velocity of the accreting material; we assume that the velocity is constant, as set by the geometry (see Equation \ref{vs}), so the range of $\curf$ represents high and low density columns.  Note that as $\curf$ decreases, larger values of $f$ are needed to produce the same amount of flux as a higher $\curf$ accretion column.
$\mdot$ can be calculated for each accretion column, with $\curf$ and $f$ known, using 
\be
\label{md}
\mdot=\frac{8\pi R_{\ast}^2}{v_s^2}\curf f.  
\en
\citep{gullbring00}.  The total $\mdot$ is a sum of the contributions from each column and from $\mdot$ we determine $L_{acc}$ using,
\be
L_{acc}=\frac{GM_{\ast}\mdot}{R_{\ast}}(1-\frac{R_{\ast}}{R_i}).
\en
This treatment does not distinguish between multiple accretion spots, each with a distinct density, or a single accretion spot with a range of densities.  The latter scenario is suggested by models produced using the Zeeman Doppler Imaging technique (ZDI) \citep{gregory11,donati08}.

In this paper, the excess which veils the photospheric absorption
lines is primarily a continuum.  The main opacity sources included in
the calculation are H bound-free and free-free, H- bound-free
and free-free, C, Si and Mg, plus additional sources described in
\citet{calvet91}.  Line blanketing is also included, using the line
list from \citet{kurucz95}.  A more thorough treatment of the spectrum
of the excess continuum including spectral lines has been presented in
\citet{dodin12} where it was found that emission lines are important
to consider when estimating veiling.  Dodin \& Lamzin showed that
veiling may be overestimated when lines are neglected in the veiling
calculation.  An important difference between the models of CG98 and
\citet{dodin12} is the structure of the region where the heated
photosphere joins the shock above it, which in both treatments is
taken at the ram pressure of the shock.  In CG98, when the heated
photosphere is joined to the postshock region the temperature rises
quickly to $\sim 10^6$ K because this region has small physical
dimensions.  Metals would become quickly ionized and few photospheric
emission cores would be expected, especially since the location of this
transition region is close to the temperature minimum.  In contrast,
since \citet{dodin12} do not have a rapid temperature rise, strong
emission lines form in this region. Additional work is needed to sort
out the differences between the two models.  For now, we acknowledge
that emission lines may affect our results, especially for the highly
veiled sources like DR Tau and RW Aur A \citep{gahm08,petrov11}, and
accretion spot sizes for those sources, in particular, may be
overestimated due to the omission of emission lines in the veiling
spectrum.

\section{Results}
\label{results}
In this analysis we include, for the first time, multiple accretion columns covering a range in energy flux; specifically, we add low $\curf$ accretion columns to high $\curf$ columns to fit both the UV and optical excesses.  Evidence for multiple accretion spots is seen in maps of the magnetosphere \citep{donati08} and also in fits of the broad He and Ca emission, which require inhomogeneity in the hot spots \citep{dodin13}.  We also compare the amount of 1 $\mu$m veiling predicted by our models with observed values.  Our spectra do not extend into the near IR, so to estimate the 1 $\mu$m veiling produced by our model we first find $J$ of our WTTS template.   We scaled $V$ by the veiling, $r_V=F_{V,veil}/F_{V,WTTS}$, and then using $V-J$ colors for the given spectral type, find the $J$ magnitude of the WTTS which reflects the veiling.  We assume that $r_J=r_{1\;\mu m}$, which is valid because the veiling continuum is constant between 0.8 and 1.4 $\mu$m \citep{fischer11,mcclure12}.  We estimate $J$ of the model by calculating the flux that would be measured assuming the transmission curve for the Two Micron All Sky Survey \citep[2MASS]{skrutskie06}.

To find the best fit of the accretion shock models to the STIS and SMARTS spectra for each source, we calculated the emission from accretion columns spanning $\curf=10^{10}-10^{12}$ erg s$^{-1}$ cm$^{-3}$, using the masses and radii from Table \ref{tabctts}.  We allowed the filling factor, $f$, to vary independently for each accretion column and then summed the flux from each column with that of the WTTS, scaled to the CTTS by the veiling at $V$ (which tells us the location of the photosphere below the excess).  The spectrum of the final model is given by;  
\be
\label{multimodel}
F_{\lambda,model}=\Sigma_i(F_{\lambda}(column, \curf_i)\times f(\curf_i))+F_{\lambda,WTTS},
\en
where $F_{\lambda,model}$ is the flux in the model, $F_{\lambda}(column, \curf_i)$ is the flux from an accretion column with energy flux $\curf_i$ and $f(\curf_i)$ is the filling factor of a column with energy flux $\curf_i$, where $\curf_i$=$10^{10},\;3\times10^{10},\;10^{11},\;3\times10^{11},\;10^{12}$ erg s$^{-1}$ cm$^{-3}$.  The values of $\curf$ we have chosen represent the typical range of values for CTTS (CG98).

When determining the best fit model, we isolated continuum regions because the accretion shock models do not attempt to reproduce line emission produced in the shock.  We calculated the $\chi_{red}^2$, as  
\be
\label{chi}
\chi^2_{red}=\frac{1}{N}\Sigma_{i=0}^N\frac{(F_{\lambda,model}-F_{\lambda,CTTS})^2}{E_{\lambda,CTTS}^2}
\en
where $N$ is the number of continuum wavelengths which contribute to the fit and $E_{\lambda,CTTS}$ is the error in the observed fluxes.  The filling factor of each contributing column in the best fit model is given in Table \ref{tabacc}; columns with $f=0$ do not contribute to the model fluxes.  Finally, we calculated $\mdot$ using Equations \ref{vs}, \ref{md} and, 
\be
\mdot=\frac{8\pi R_{\ast}^2}{v_s^2}\times\Sigma_i(\curf_i \times f(\curf_i))
\en
Estimated $\mdot$s are listed in Table \ref{tabacc}, along with the sum of the filling factors of all columns.  

\begin{figure}
\plotone{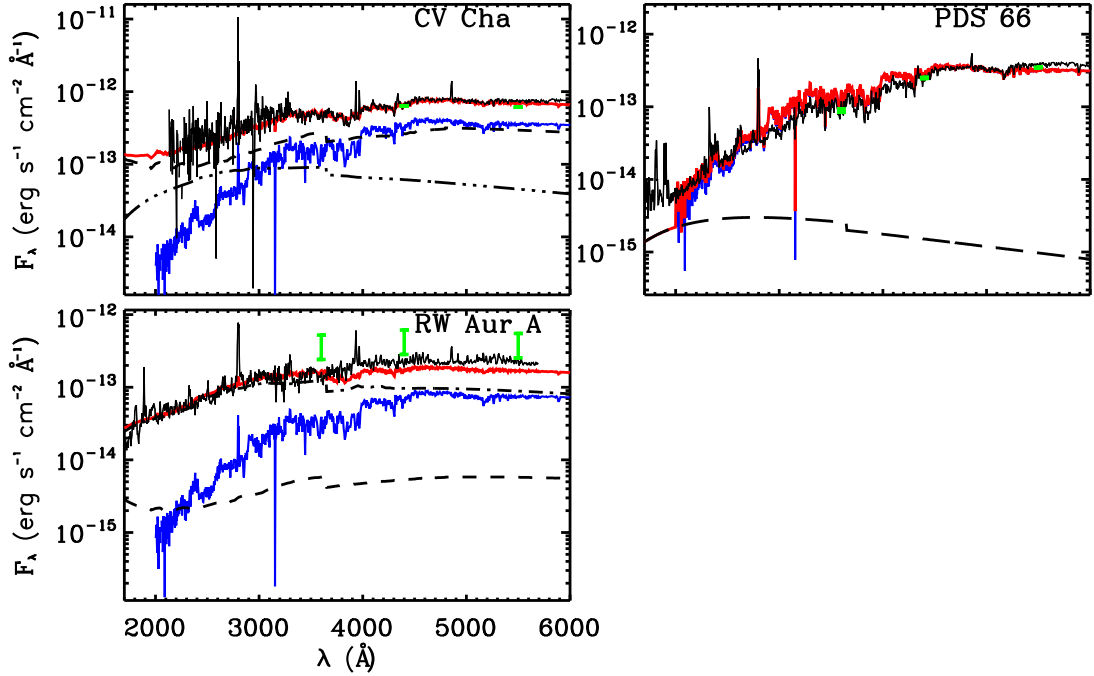}
\caption{Spectra of late G and early K spectral type CTTS in the DAO sample: CV Cha, PDS 66 and RW Aur A.  We fit the STIS NUV and optical spectra with emission from accretion columns plus a WTTS spectrum.  In each panel the black spectrum is the CTTS and the blue spectrum is the WTTS, LkCa 19.  The broken black lines represent accretion shock models with different $\curf$ values, defined as in Figure \ref{shocks}.  The red line is the best model fit to the data, adding the emission from the different shock models to the WTTS spectrum.  The green error bars indicate non-simultaneous photometry, representing the range observed in multi-epoch observations when available.  Although PDS 66 does not have an NUV excess, examination of the H$\alpha$ emission line reveals that it is broad, and therefore accretion is likely ongoing, though at low levels.}
\label{lkca19}
\end{figure}

\begin{figure}[htp]
\plotone{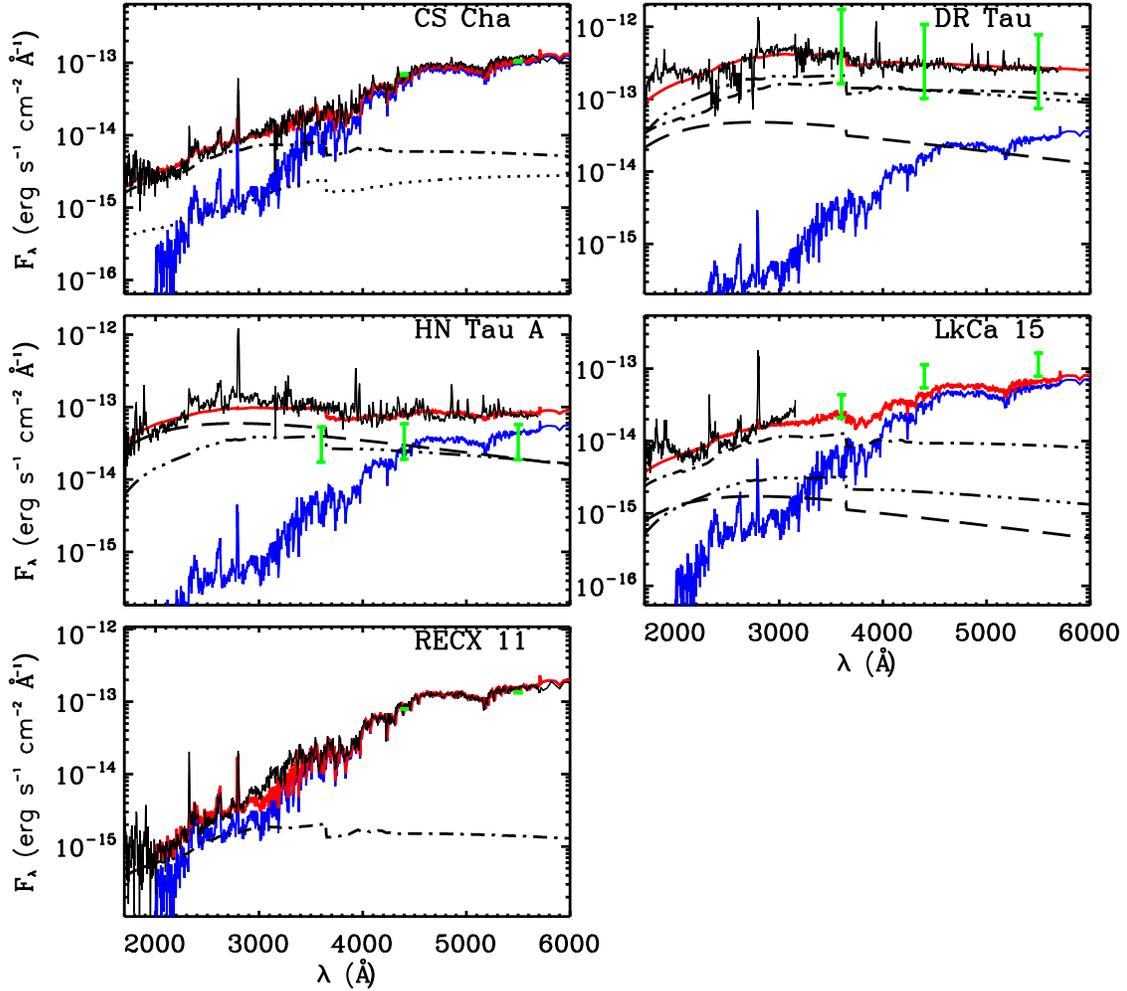}
\caption{Spectra of mid K CTTS in the DAO sample: CS Cha, DR Tau, HN Tau A, LkCa 15 and RECX 11.  Lines are defined as in Figure \ref{lkca19}.  The adopted WTTS in each panel is RECX 1.  RECX 11 does not have a significant NUV excess, but it was shown in \citet{ingleby11b} that it is still accreting.}
\label{recx1}
\end{figure}

\begin{figure}[htp]
\plotone{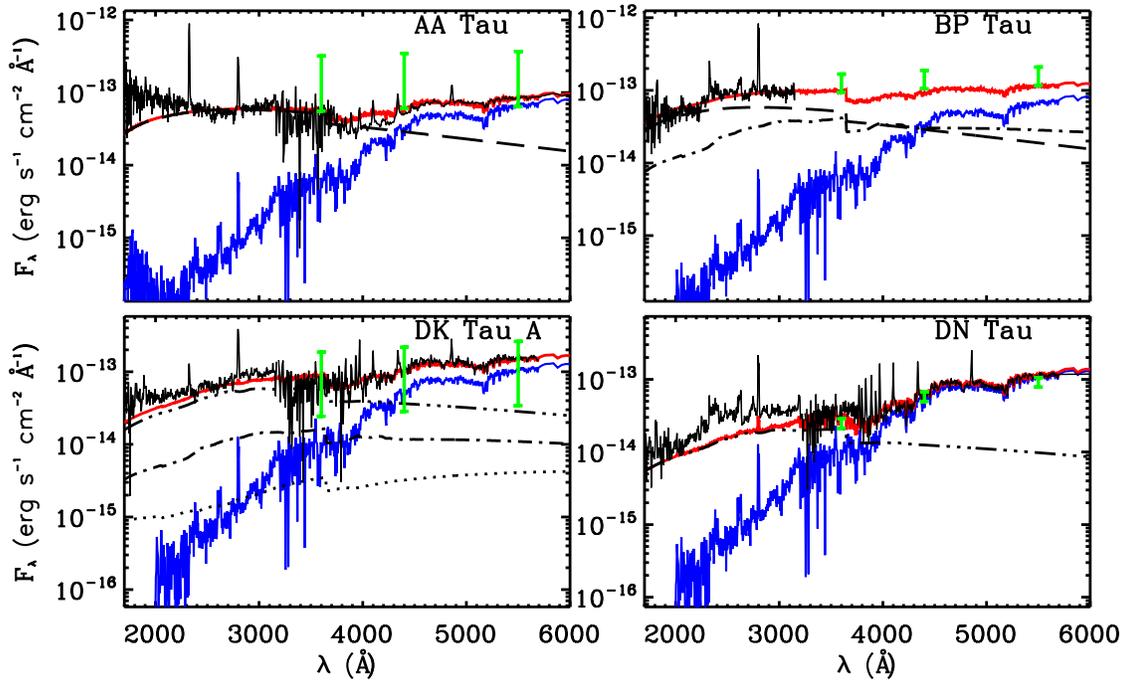}
\caption{Spectra of late K and early M CTTS in the DAO sample: AA Tau, BP Tau, DK Tau A, DN Tau.  Lines are defined as in Figure \ref{lkca19}.  The adopted WTTS in each panel is HBC 427.}
\label{hbc427a}
\end{figure}

\begin{figure}[htp]
\plotone{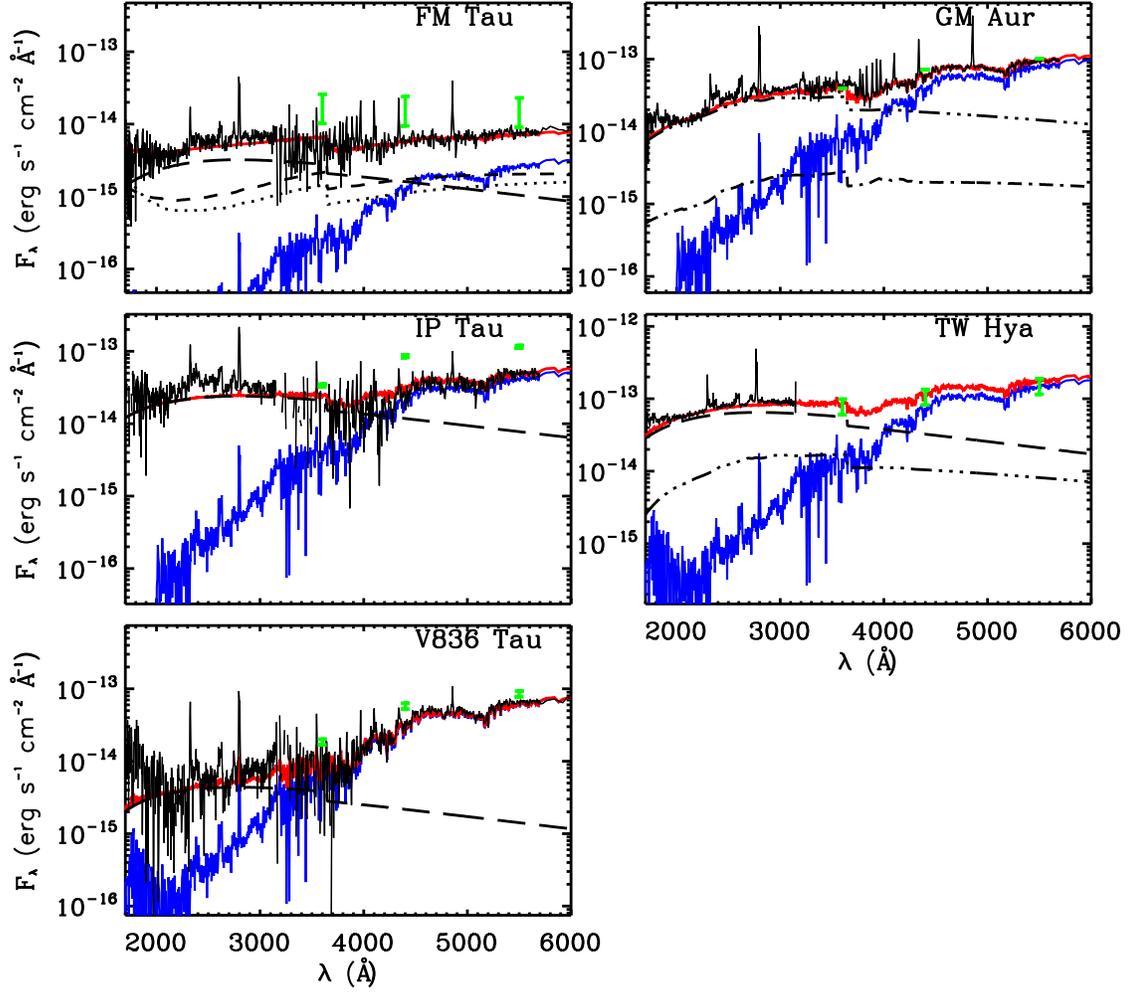}
\caption{Spectra of late K and early M CTTS in the DAO sample continued: FM Tau, GM Aur, IP Tau, TW Hya and V836 Tau.  Lines are defined as in Figure \ref{lkca19}.  The adopted WTTS in each panel is HBC 427.}
\label{hbc427b}
\end{figure}

\begin{figure}[htp]
\plotone{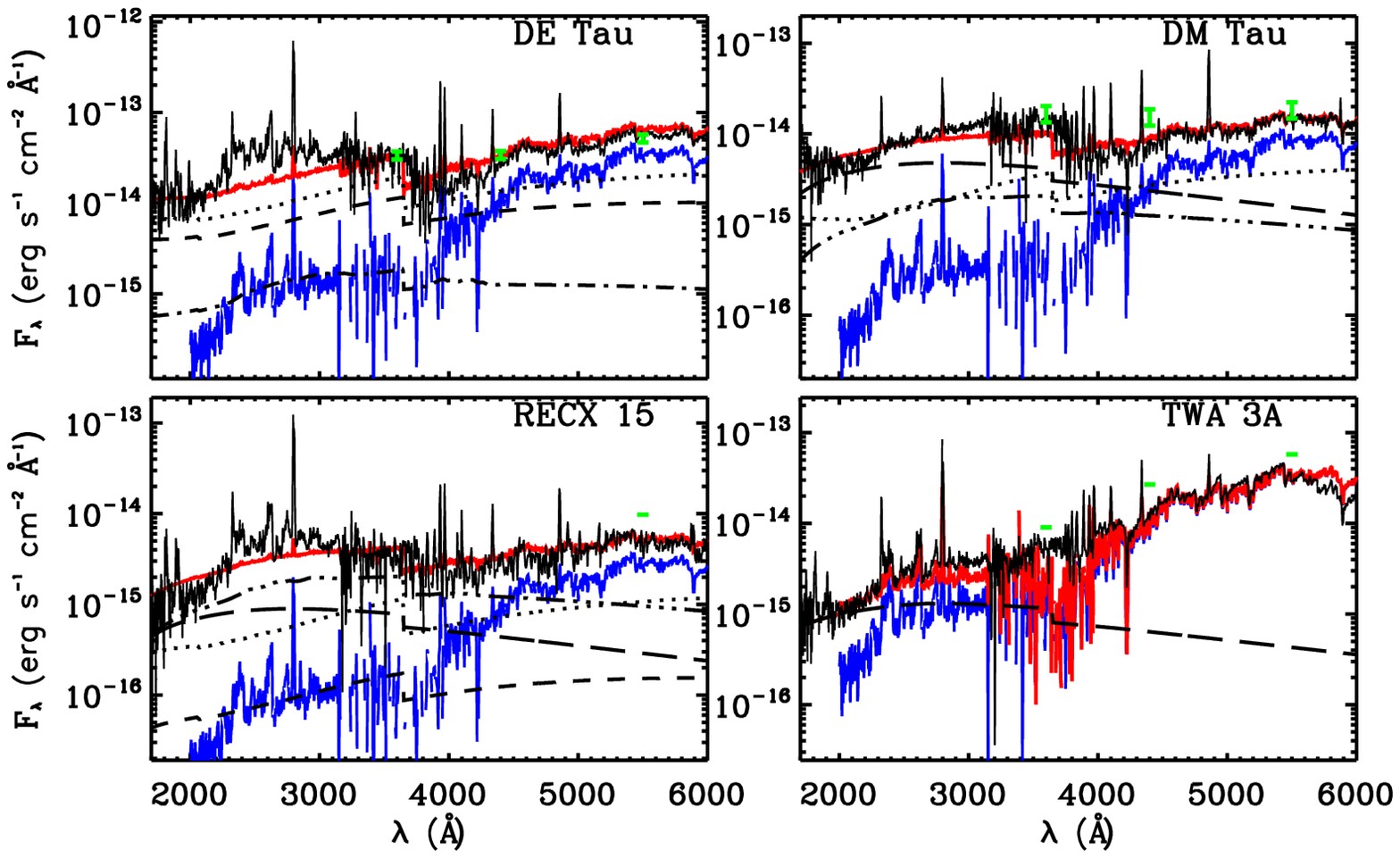}
\caption{Spectra of mid M CTTS in the DAO sample: DE Tau, DM Tau, RECX 15 and TWA 3a.  Lines are defined as in Figure \ref{lkca19}.  The adopted WTTS in each panel is TWA 7.}
\label{twa7}
\end{figure}

The best fit of the accretion shock models to the de-reddened NUV and optical spectra, corresponding to the stellar and accretion properties listed in Tables \ref{tabctts} and \ref{tabacc}, are shown in Figures \ref{lkca19} -- \ref{twa7}.  Overall, we found good fits between the models and the observed spectra, with a few exceptions.  For the later spectral types, especially the M stars, the models did not reproduce a rise in the observed spectra between 2000 and 3000 {\AA} (see, for example, RECX 15 and DE Tau in Figure \ref{twa7}).  This spectral region is populated with Fe emission lines, which are unresolved at the STIS resolution.   The Fe lines may be produced in accretion related processes \citep{herczeg05,petrov11}; however, they are also observed in the WTTS templates, though weaker, they appear to have a chromospheric component.  Fe lines become more apparent at later spectral types, as the photospheric emission in the UV decreases.  Finally, we do not attempt to fit the FUV spectrum ($<$1700 {\AA}) which has contributions from \h2 in the disk \citep{ingleby09} that are not included in the accretion shock model.  

Figures \ref{lkca19} -- \ref{twa7} also include non-simultaneous photometry from the literature, de-reddened using the $A_V$s from Table \ref{tabctts}.  When available, we used the range of optical photometry from \citet{herbst94} and this range is shown as the green error bars.  The photometry for the remaining CTTS came from the following sources; CV Cha \citep{lawson96}, GM Aur, IP Tau, LkCa 15, TW Hya and V836 Tau \citep{kenyon95}, PDS 66 \citep{batalha98,cortes09}, RECX 11 and RECX 15 \citep{sicilia09} and TWA 3a \citep{gregorio92}.  The models shown do not attempt to fit the photometry because it is not simultaneous, but we use the photometry to look for  evidence of high amplitude variability.  Most of the photometry agrees with the STIS and SMARTS spectra and is therefore well fit by the models.  For a number of the sources, RW Aur A, FM Tau, IP Tau, V836 Tau, RECX 15 and TWA 3A, the STIS spectra are slightly lower than expected from the photometry.  Of these sources with mis-matched photometry and spectra, with the exceptions of RW Aur A and FM Tau, photometry for only one epoch is available, so the difference may be due to variability.  The DAO NUV and optical spectra of HN Tau A are significantly higher than the range of photometry from \citet{herbst94}; however, \citet{grankin07} observed that the brightness of HN Tau A was steadily decreasing during the 10 years between 1985 and 1995.  It may be that \citet{herbst94} observed HN Tau A during a dim period and the source has since brightened; such changes in brightness are observed in other CTTS, so it is not unexpected \citep{grankin07}.

In Table \ref{tabacc} we also give the values of $r_Y$ predicted by our models.  Comparing the model $r_Y$ to those from the literature or our observations (Table \ref{tabctts}), the model values for all but four sources are within $\pm$0.1 of the observed $r_Y$, indicating that near IR veiling naturally occurs from a multi-column accretion shock model.  Sources with large $r_Y$ are fit by models with large filling factors for the low $\curf$ columns.  The four sources which do not agree are AA Tau, DR Tau, DK Tau A and HN Tau A.  For DK Tau A, we were unable to reproduce the 1 $\mu$m veiling, likely because of variability.  In \citet{edwards06}, the veiling at $V$ is equal to the veiling at 1 $\mu$m, requiring an abnormally red accretion spectrum, which is not supported by our NUV and optical observations.  The $V$ and 1 $\mu$m veiling measurements were not simultaneous, so it is possible that $\mdot$ was higher when the IR observations were obtained.  HN Tau A also appears to be a remarkably variable object, as mentioned above, so it is possible that variability is responsible for not being able to fit the non-simultaneous optical and near IR veiling data.  Veiling is very difficult to measure for DR Tau because the excess continuum is so strong that many absorption lines are completely filled in; therefore, we expect large error bars on the near IR veiling.

\begin{figure}[htp]
\plottwo{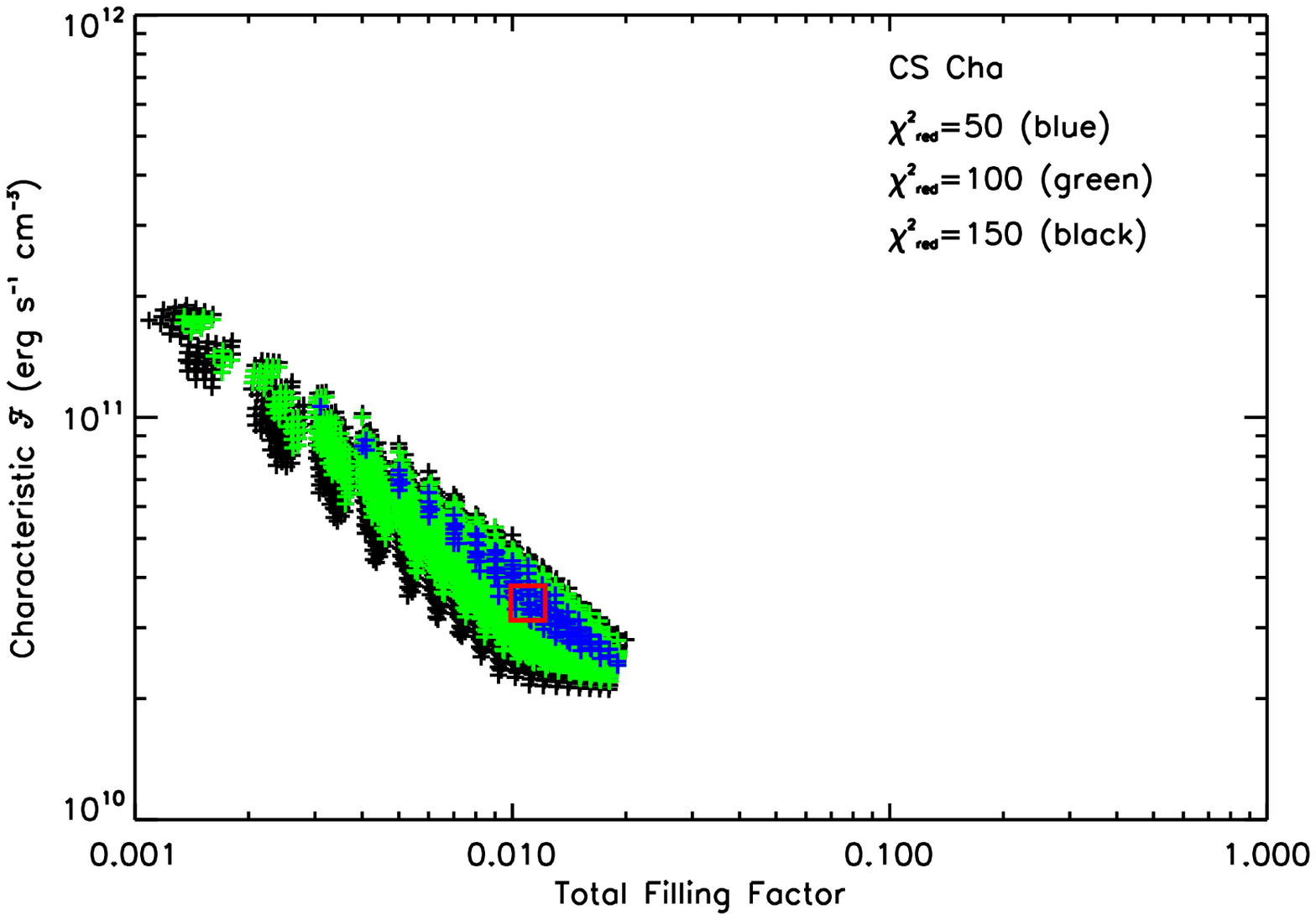}{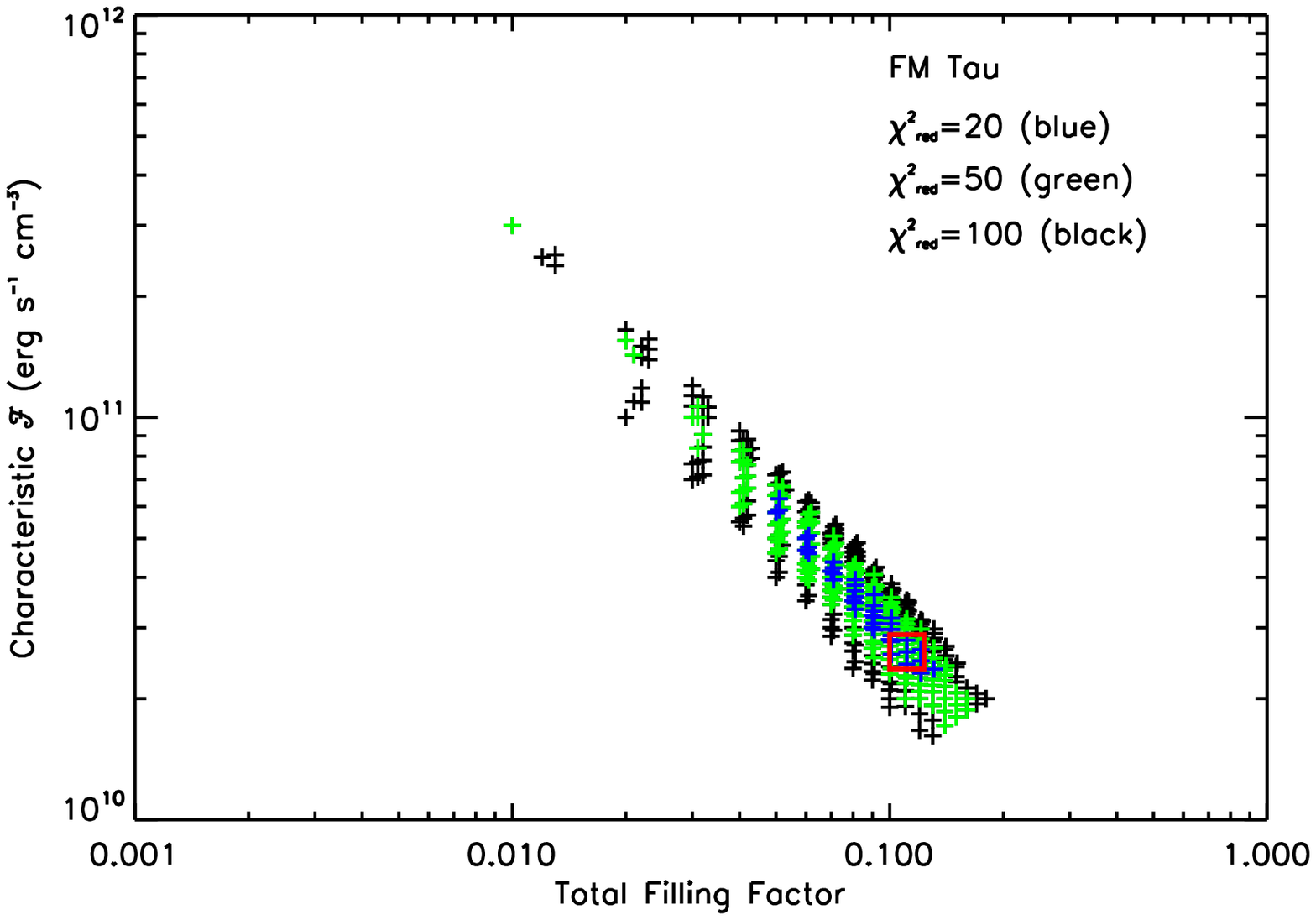}
\caption{Degeneracy in characteristic $\curf$ and total filling factor f.  The color of each point represents the value of $\chi^2_{red}$ calculated from Equation \ref{chi}, assuming a model with a given total filling factor and characteristic $\curf$ (the average $\curf$ of the contributing columns weighted by the filling factor) for CS Cha (left) and FM Tau (right).  The plot range represents a typical parameter space over which the accretion shock models were calculated.  The red box shows the point where $\chi^2_{red}$ is at a minimum.  }
\label{cont}
\end{figure}

The models in Figures  \ref{lkca19} -- \ref{twa7} represent the best fit to the UV and optical data; however there is some degeneracy in $\curf$ and $f$.  There is a range in $\curf$ and $f$ which will produce a shock spectrum that results in a comparable $\chi^2_{red}$ fit to the data.  Figure \ref{cont} shows this range for CS Cha and FM Tau, where the colored points represent values of $\chi^2_{red}$ for models with different total filling factors and characteristic $\curf$ values.  The characteristic $\curf$ value is the average $\curf$ of the contributing columns, weighted by the filling factor of each column.  Figure \ref{cont} reveals that there is a range of roughly one order of magnitude in both $\curf$ and $f$ for which models will result in fits with $\chi^2_{red}$ $<2\times$ the best fit model.  While interesting for the shock geometry, this degeneracy has little affect on the $\mdot$'s because each model is accurately measuring the total excess from which the accretion luminosity is measured.

An important result is that, by including the low $\curf$ columns, we find significantly higher filling factors than would be needed to fit the UV excess alone (Table \ref{tabacc}).  CG98 fit optical spectra of Taurus CTTS using a single accretion column model.  For seven out of nine of our sources which overlap with CG98, we find higher $f$ values.  The characteristic $\curf$ of the contributing accretion columns is typically lower than that found by CG98,  because of the addition of low $\curf$ accretion columns which have large filling factors.  However, both DN Tau and AA Tau have larger filling factors \emph{and} higher $\curf$ values in CG98 than in our analysis.  DN Tau was fit by an accretion column characterized by $log\; \curf=10.5$ and $f=0.5\%$ in CG98; however, their fit to the observed spectrum cannot explain the blue excess very well.  Our model includes a high $\curf$ column along with the low $\curf$ column, better fitting both the blue and red excesses and skewing the characteristic $\curf$ to a higher value.  We mentioned above that our model for AA Tau did not reproduce near IR veiling measurements.  There may be variability in the cool accretion component, where it was less prevalent during the DAO observations.  Were we to increase the contribution from low $\curf$ columns to fit the 1 $\mu$m veiling, it would bring the characteristic $\curf$ to a lower value and the $f$ to a larger value than found in CG98.  Given that models of the magnetosphere predict large $f$ \citep{long11}, multi-component accretion columns are a better representation of the physical geometry of the system and now reproduce the flux distribution as well.

\begin{figure}[htp]
\plotone{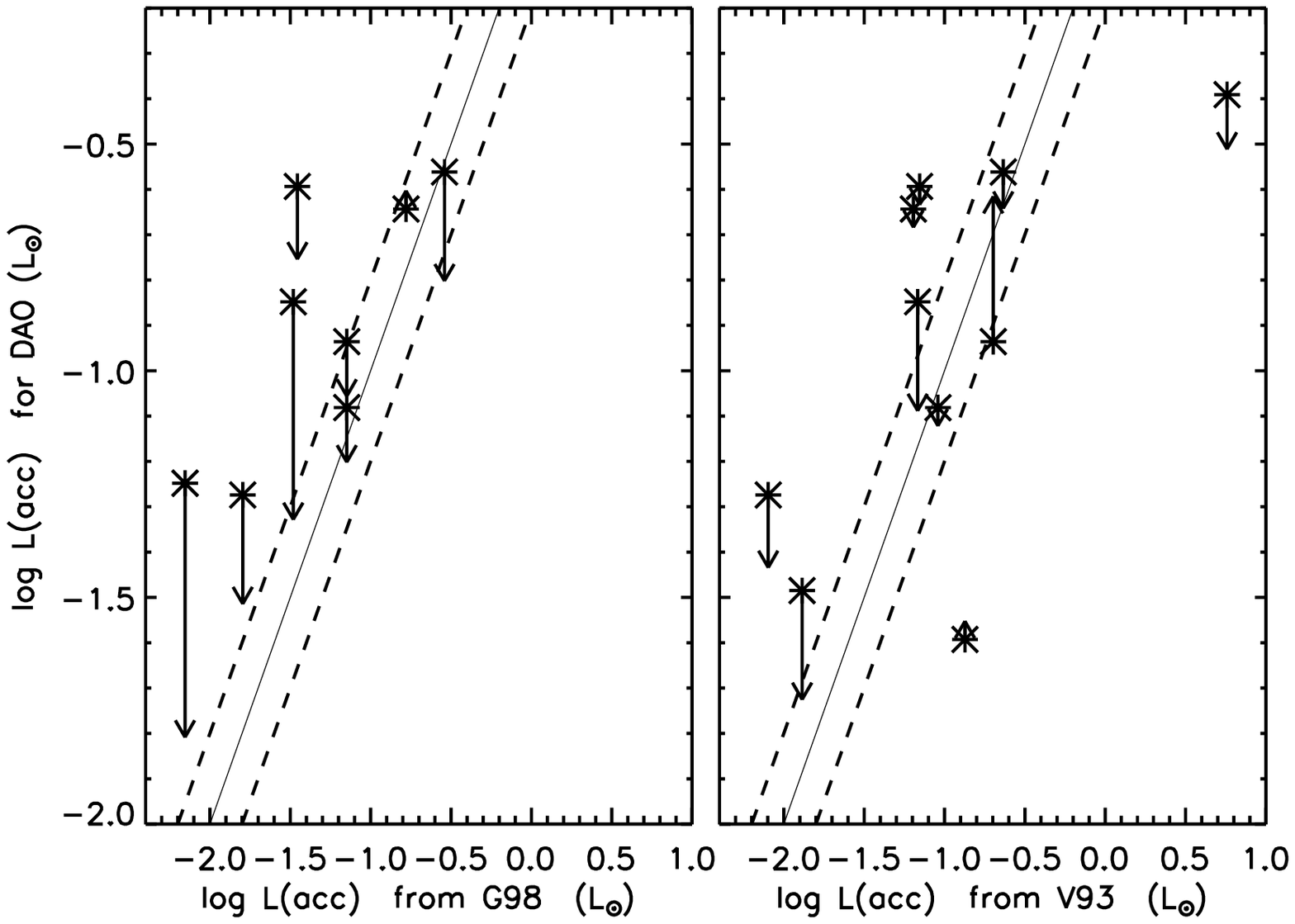}
\caption{DAO accretion luminosities versus values from the literature.  We compare our estimates of $L_{acc}$ to those from \citet{gullbring98} on the left and \citet{valenti93} on the right.  The solid line shows where sources would fall if the accretion luminosities agreed.  The dashed lines represent a $\pm$30\% difference in the values of $L_{acc}$.  Asterisks represent accretion luminosities calculated assuming the value of $A_V$ listed in Table \ref{tabctts} while the arrows show how $L_{acc}$ would change if we use the $A_V$s from \citet{gullbring98} or \citet{valenti93}.}
\label{mdotmdot}
\end{figure}

We compare the accretion properties calculated here with those from two previous studies of young stars in Taurus in Figure \ref{mdotmdot} and Table \ref{tabacclit}.  \citet{valenti93}, hereafter V93, fit optical spectra with coverage from 3400 to 5000 {\AA} of Taurus CTTS using a hydrogen slab model and a WTTS template.  The slab was characterized by the temperature, number density, thickness and coverage on the stellar surface.  \citet{gullbring98}, hereafter G98, used similar assumptions to those of V93 and therefore, their accretion rates are in good agreement. G98 notes that the major differences between their $\mdot$s and the V93 $\mdot$s come from choices of $A_V$ and the evolutionary tracks used to determine $M_{\ast}$.  Different $A_V$s change the stellar luminosity, and therefore radius, so the value of $M_{\ast}/R_{\ast}$ will be affected by the extinction estimate.  To avoid the additional variables of mass and radius in the analysis, we compare the accretion luminosities from each study instead of $\mdot$.

In Figure \ref{mdotmdot}, left, we plot our values of $L_{acc}$ compared to those of G98 and find that our estimates tend to be higher.  The values of $L_{acc}$ we calculate here (shown by the asterisks) have less than a 30\% difference from those of the G98 for half of the sample.  After correcting our accretion luminosities by the difference in $A_V$, shown by the arrows, we find that all but one are in good agreement.    The outlier is HN Tau A, where our $\mdot$ is over one magnitude larger than G98.  This is likely due to real variability; we observed that the STIS fluxes are higher than the range in optical photometry observed by \citet{herbst94}.    It appears that two factors in our analysis, the excess flux decrement in the UV due to the WTTS template and the additional excess flux in the red due to the low $\curf$ column, cancel each other out.  Therefore, we find similar values of $L_{acc}$, after accounting for differences in $A_V$, to those of G98.  We chose to use the extinction estimates of \citet{furlan09,furlan11} who used IR colors to estimate the amount of extinction, avoiding spectral regions where the shock contribution is highest.  \citet{furlan09,furlan11} also covered all of our sources in the Taurus and Chamaeleon I regions, allowing us to use $A_V$'s which were derived consistently for each source. 

When comparing the DAO accretion luminosities to those of V93, our estimates are high for three sources; DK Tau A, HN Tau A and DN Tau and low for two; FM Tau and RW Aur A.  V93 obtained optical spectra from the UV Schmidt spectrometer at Lick Observatory and provided an atlas of the observations making direct comparison of the optical fluxes possible.  For DK Tau A, HN Tau A and DN Tau, the optical fluxes are lower in V93 than in our observations while for RW Aur A and FM Tau they are higher, consistent with the discrepancies in $L_{acc}$.  These variations in the observed fluxes may be intrinsic variability but the difficulty of flux calibrating ground based slit spectra may also contribute.  In particular for FM Tau, V93 notes that their slit loss correction was greater than 25\%, indicating that it was observed at a high zenith angle or in poor seeing, making flux calibration more uncertain.

\subsection{Hidden Accretion Emission}
\label{hidden}

\begin{figure}[htp]
\plotone{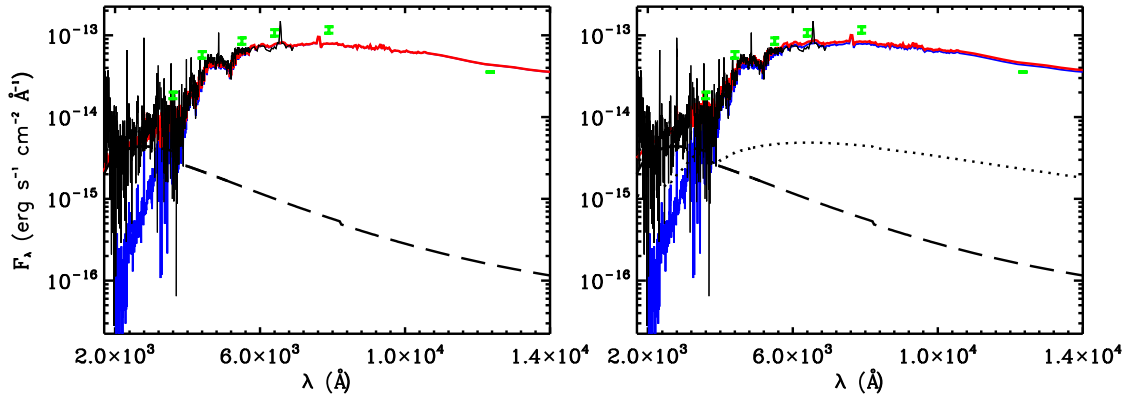}
\caption{Range of possible accretion shock models for V836 Tau.  In each panel the lines are defined as in Figure \ref{lkca19}.  The left panel shows the best fit model of the accretion shock to the optical spectra.  The right panel shows the fit if we assume that some accretion flux is hidden by the stellar photosphere.  The upper limit on the luminosity of the low $\curf$ accretion column (dotted line), constrained by the lack of veiling at 1 $\mu$m, is equal to that of the column which fits the NUV excess (long dashed line), doubling $\mdot$.}
\label{v836tau}
\end{figure}

Not all sources require a low energy column to fit the UV and optical excesses in our analysis; however, there could be emission from cold columns hidden by the photospheric flux. If these columns were present, they would increase our estimates of the mass accretion rates.  Here, we determine how much hidden flux may be present for sources with no detectable red excess.  In Figure \ref{v836tau} we show two model fits to the spectrum of V836 Tau, which has no observed veiling at 1 $\mu$m \citep{edwards06}.  In the left panel, the best fit model is produced with only a high $\curf$ column, no additional accretion columns are needed (Table \ref{tabacc}).  In the right panel, we assume that a low $\curf$ column exists, but the emission from the column is not detectable above the stellar component.  For V836 Tau we find that the contribution to the accretion emission that may be hidden below the stellar emission could be equal to that in the high $\curf$ column, doubling the estimated accretion rate.  This new model meets the constraint of having limited veiling at 1 $\mu$m, with $r_{1\;\mu m}<$ 0.1, within the errors of veiling estimates.

We perform this analysis for all sources which had $f=0$ for the low $\curf$ columns in the initial fit, with the constraints that the NUV and optical fluxes are not overestimated by the new models, and that the veiling at 1 $\mu$m remains within $\pm$0.1 of the observed 1 $\mu$m veiling.  Using this method, we calculate upper limits on $\mdot$ and the filling factor, listed in Table \ref{tabacc2}, where the differences in the two filling factors comes from increasing $f(1\times10^{10}$) from its value in Table \ref{tabacc}.  These additional low $\curf$ accretion columns are important because they increase the expected $\mdot$ along with the filling factors.  For the example shown in Figure \ref{v836tau}, the accretion rate would double if we include the maximum shock emission that could be hidden by the photosphere, but the filling factor may be up to 100 times higher.  The increase of the filling factor is much higher than $\mdot$ because there is less mass per unit area in the low $\curf$ columns than in the high $\curf$ columns.  This would increase the filling factor from $<$0.1\% to a few percent, better in line with the models of the accretion footprints produced by complex magnetic field geometries \citep{long11} and the distribution of excess accretion related emission regions now being found from magnetic mapping studies \citep{donati11}.  The addition of a hidden accretion column is increasingly significant for sources with a lower accretion rate, where the flux that may be hidden by the star becomes comparable to the excess observed in the UV.  

\section{Correlations with Accretion Indicators}
\label{correlations}
Measuring the UV excess is ideal for calculating all but the lowest $\mdot$s; however, UV observations are difficult to obtain.  For this reason it is common to use an indicator of accretion easily accessible in the optical or near-IR which has been calibrated with $\mdot$s calculated for the few sources with UV observations.  We assume that broad optical line emission, primarily in lines of hydrogen and calcium, originates in the material free-falling onto the star in the accretion flows supported by the fact that the emission lines are well reproduced by models which assume this geometry \citep{muzerolle98,muzerolle01}.  Recently, \citet{dupree12} suggested that the lines are instead formed in the turbulent region directly below the shock.     However, 
\citet{calvet81} found that even a chromosphere
covering the entire stellar surface did not have enough
emitting volume to explain the lower Balmer lines in CTTS;
the shock region also has a small emitting volume
and so it is unlikely to explain the observed emission in these
lines.  

Both the equivalent width of H$\alpha$,  EW(H$\alpha$), and the width of H$\alpha$ at 10\% of the maximum flux are often used as  accretion proxies \citep{white03,natta04}.  The width of H$\alpha$ at 10\% is expected to be a better indicator of accretion, since the wings of the line trace the fast moving material in the accretion flows \citep{muzerolle01}.  Also, the EW(H$\alpha$) saturates for the highest accretors, or sources with the largest veiling \citep{muzerolle98}.    Correlations between $\mdot$ and H$\alpha$ 10\% width have been widely used; however there is considerable scatter because H$\alpha$ and the UV excess have not been measured simultaneously.  The dataset presented here, with UV and H$\alpha$ observations separated by less than one to a few days, would be ideal for this analysis; however, the resolution of the SMARTS optical spectra is too low to accurately measure the 10\% width of H$\alpha$.  Even observing the wings of H$\alpha$ may not be the best method for determining $\mdot$, as \citet{costigan12} showed that variability in the wings exceeded variability observed in other tracers.

\begin{figure}[htp]
 \plotone{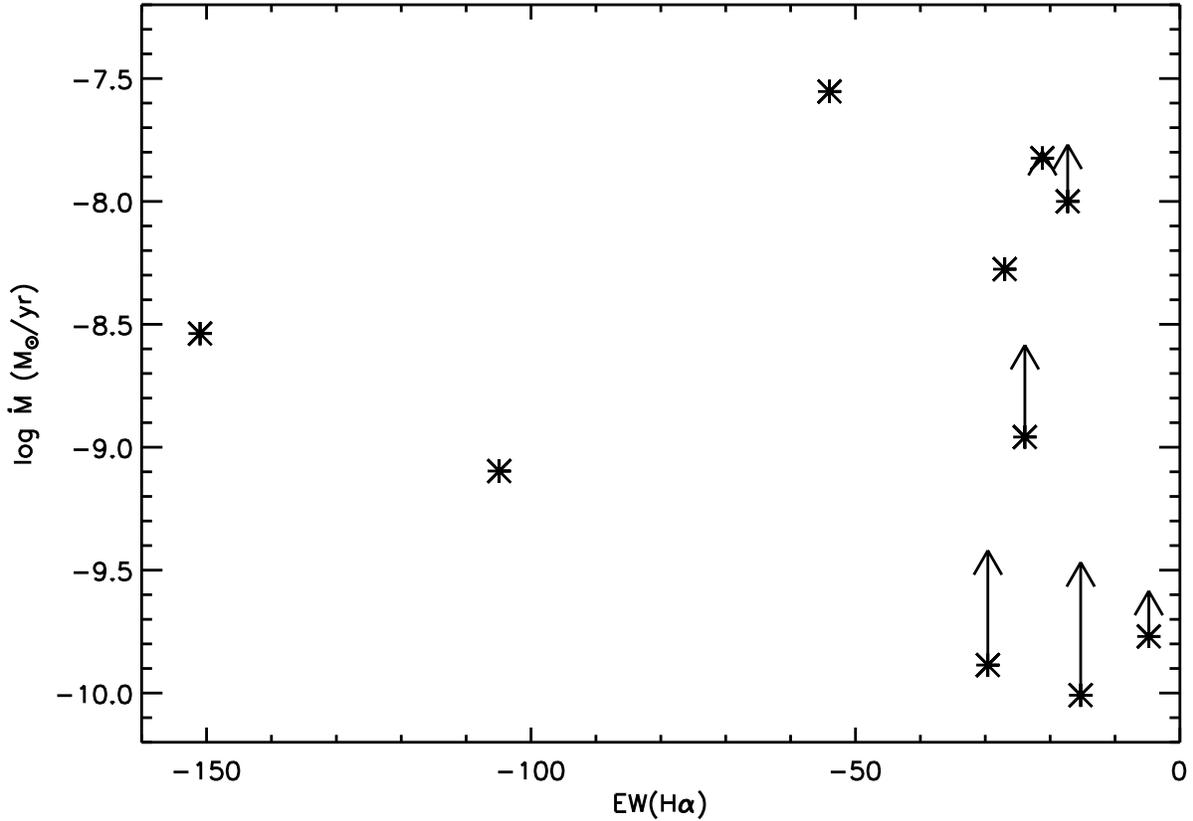}
\caption[$\mdot$ versus EW(H$\alpha$)]{$\mdot$ versus EW(H$\alpha$).  We compare our calculated $\mdot$s given in Table \ref{tabacc} to the EW(H$\alpha$) measured from the nearly simultaneous SMARTS spectra.  The asterisks show the initial fits to the UV and optical excesses.  Arrows show the range of $\mdot$ for sources which were assumed to have a hidden cool accretion component, with the point of the arrow representing the upper limit on $\mdot$.  We see no correlation between the EW(H$\alpha$) and $\mdot$.}
\label{ew}
\end{figure}

\begin{figure}[htp]
\plotone{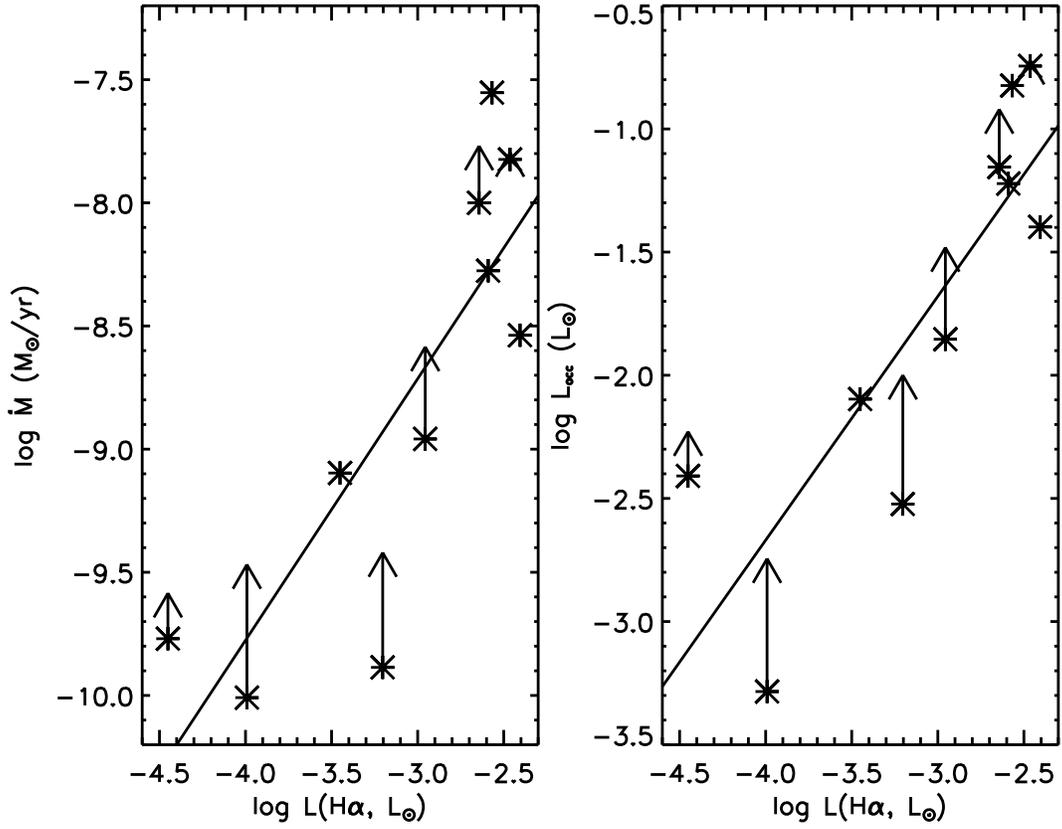}
\caption[$\mdot$ and $L_{acc}$ versus H$\alpha$ line luminosity]{$\mdot$ and $L_{acc}$ versus H$\alpha$ line luminosity.  The luminosity of H$\alpha$, measured in the low resolution SMARTS spectra, is correlated with the accretion properties. Asterisks and arrows are defined as in Figure \ref{ew}.  Least square fits to the data are given in Equations \ref{mdothalpha} and \ref{lacchalpha}.}
\label{halpha}
\end{figure}

In Figure \ref{ew}, $\mdot$ and EW(H$\alpha$) are plotted for the sample of our sources which had nearly simultaneous SMARTS optical and UV observations.  There is no clear relation between EW(H$\alpha$) and $\mdot$.  Two sources, DM Tau and RECX 15, have relatively low $\mdot$ but significant EW(H$\alpha$), showing that it is not a good tracer of $\mdot$.  However, \citet{herczeg08} and \citet{manara12} found correlations between the luminosity of H$\alpha$ and $\mdot$ or $L_{acc}$.  \citet{herczeg08} determine $L_{acc}$ using optical spectra with coverage down to the atmospheric limit in the blue while \citet{manara12} observed the $U$ excess (to estimate $L_{acc}$) and H$\alpha$ with the \emph{HST} Wide-Field Planetary Camera 2 (WFPC2).  Our analysis is the next step in defining these correlations because we have long wavelength spectral coverage, in particular the STIS spectra provide the shape of the excess in the UV.  We can measure the flux of the H$\alpha$ line in the SMARTS data and $\rm{L_{H\alpha}}$ is then calculated using distances from Table \ref{tabctts}.  In Figure \ref{halpha} we see a correlation between H$\alpha$ luminosity and $\mdot$ (or $L_{acc}$) with a Pearson correlation coefficient of 0.9.  The least-square fit to the data in Figure \ref{halpha} yields
\be
\label{mdothalpha}
\rm{log(\mdot)=1.1(\pm0.3)log(L_{H\alpha})-5.5(\pm 0.8)}
\en
\be
\label{lacchalpha}
\rm{log(L_{acc})=1.0(\pm0.2)log(L_{H\alpha})+1.3(\pm 0.7)}
\en
For Equations \ref{mdothalpha}--\ref{laccmgii},  $\mdot$ is in units of $\msunyr$ and all luminosities are in units of $\lsun$.  Equation \ref{lacchalpha} is in good agreement with the fits found in \citet{herczeg08} and differs only slightly from that of \citet{manara12}, who found a y-intercept of 2.6$\pm$0.1 as opposed to our value of 1.3$\pm$0.7.

\begin{figure}[htp]
\plotone{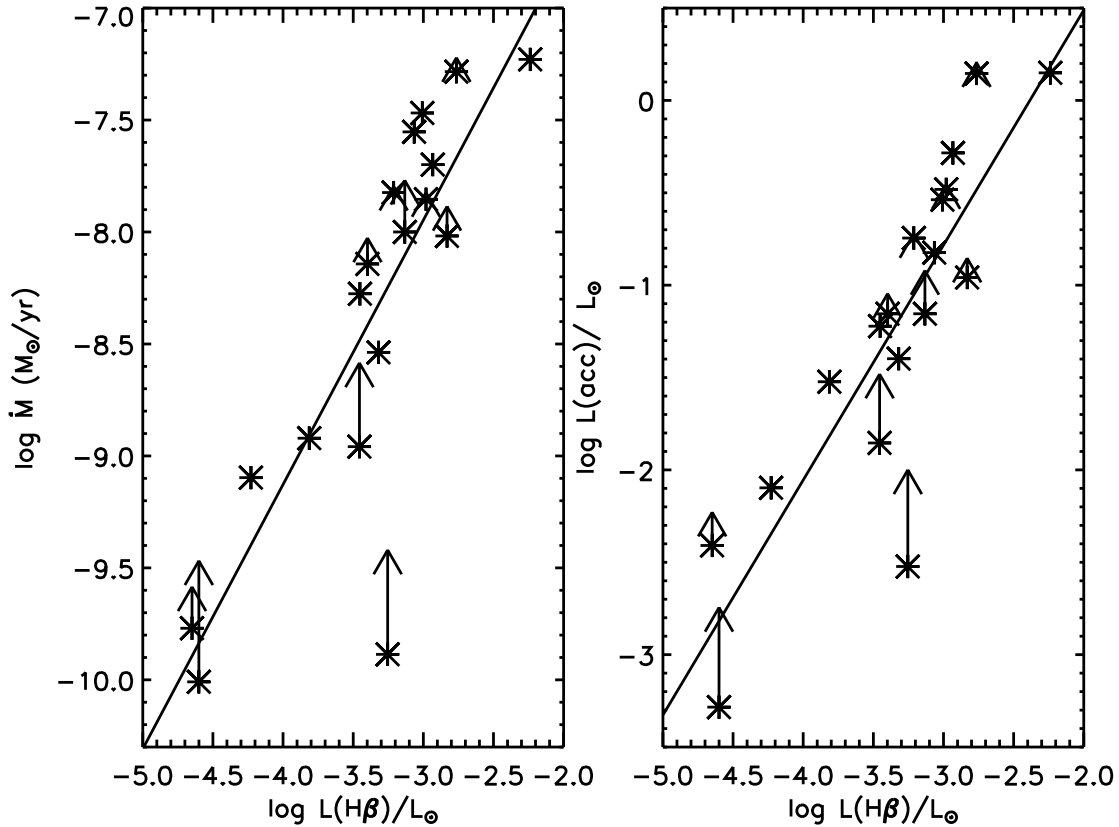}
\caption[$\mdot$ and $L_{acc}$ versus H$\beta$ line luminosity]{$\mdot$ and $L_{acc}$ versus H$\beta$ line luminosity.  Asterisks and arrows are defined as in Figure \ref{ew}.  We measured the luminosity of H$\beta$ from the simultaneous STIS spectra.  H$\beta$ clearly traces the accretion properties of the source.  Fits to the data are given in Equations \ref{hbetamdot} and \ref{hbetalacc}.}
\label{fighbeta}
\end{figure}

H$\beta$ is less often used to trace accretion than H$\alpha$ but has been shown to correlate with $\mdot$ \citep{muzerolle01,herczeg08,fang09}.  We compare the luminosities in H$\beta$ ($\rm{L_{H\beta}}$) measured from the simultaneous STIS optical spectra to both $\mdot$ and $L_{acc}$ in Figure \ref{fighbeta}.  We find strong correlations between the line luminosity and the accretion rates and luminosities, both with correlation coefficients $\sim$0.9.  The lines shown in Figure \ref{fighbeta} which describe the trends are given by the following equations;
\be
\label{hbetamdot}
\rm{log(\mdot)=0.9(\pm0.1)log(L_{H\beta})-5.1(\pm 0.5)}
\en
\be
\label{hbetalacc}
\rm{log(L_{acc})=1.0(\pm0.1)log(L_{H\beta})+2.4(\pm 0.5)}.
\en
Again, our relation between $L_{acc}$ and $L_{H\beta}$ agrees with that found by \citet{herczeg08}.

\begin{figure}[htp]
 \plotone{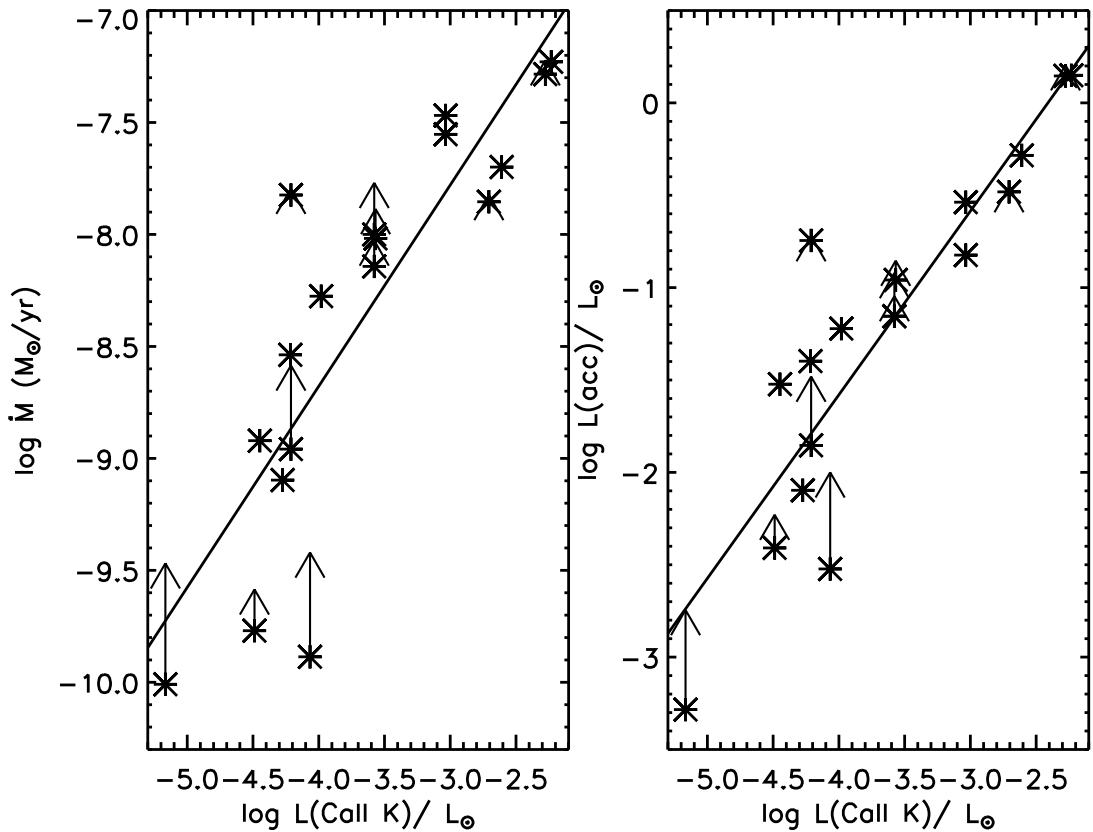}
\caption[$\mdot$ and $L_{acc}$ versus the Ca II K line luminosity]{$\mdot$ and $L_{acc}$ versus the Ca II K line luminosity.  Asterisks and arrows are defined as in Figure \ref{ew}.  We measured the luminosity of the Ca II K line from the simultaneous STIS spectra.  Like the infrared triplet lines of Ca II, the K line luminosity correlates with the accretion properties.  Fits to the data are given in Equations \ref{mdotcaii} and \ref{lacccaii}.}
\label{figcaiik}
\end{figure}

Another commonly used tracer of accretion is the Ca II infrared triplet line emission \citep{mohanty05,rigliaco11}.  Our SMARTS spectra did not cover the infrared triplet lines of Ca II, but in our STIS optical data, we observe the Ca II H and K lines.  The H line at $\lambda$3969 {\AA} is blended with the $\lambda$3970 {\AA} H$\epsilon$ line at the resolution of STIS, so we only consider the K line which is free of contamination.  Both the H and K lines are also used as tracers of chromospheric activity \citep[and references therein]{mamajek08}.  We measured the luminosity of the K line in our sample of WTTS and found that the maximum value of log $L(Ca\; II\; K)=-4.3\;\lsun$.  For CTTS with similar Ca II K luminosities, there will be some contamination in the line due to chromospheric activity. Similar to the H$\alpha$ and H$\beta$ line luminosities, the Ca II K line at $\lambda$3934 {\AA} appears to have an origin in accretion related processes, as there is a clear trend between the Ca II K luminosity ($\rm{L_{CaII\; K}}$) and $\mdot$ or $L_{acc}$ (with correlation coefficients of 0.8 in each case).  We show the trends in Figure \ref{figcaiik} and give the relations in the following;
\be
\label{mdotcaii}
\rm{log(\mdot)=0.9(\pm0.2)log(L_{CaII\; K})-5.1(\pm 0.7)}
\en
\be
\label{lacccaii}
\rm{log(L_{acc})=1.0(\pm0.2)log(L_{Ca II\; K})+2.2(\pm 0.7)}
\en
Equation \ref{lacccaii} is once again in good agreement with the relation between $L_{acc}$ and $L_{Ca II\; K}$ produced in \citet{herczeg08}.

\begin{figure}
 \plotone{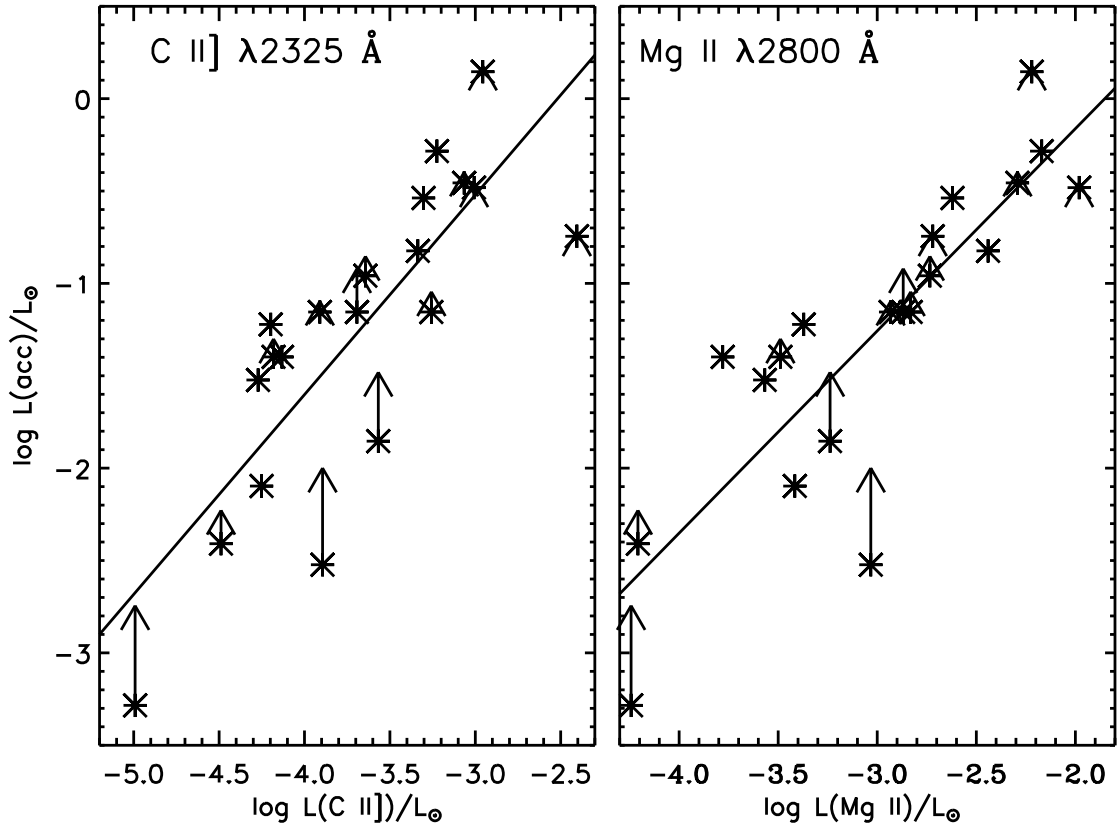}
\caption[Accretion luminosity versus luminosity of NUV emission lines]{Accretion luminosity versus luminosity of NUV emission lines.  We compare the luminosity of C II] $\lambda$2325 {\AA} (left panel) and Mg II $\lambda$2800 {\AA} (right panel) to the range of accretion luminosities calculated from the $\mdot$s in Table \ref{tabacc}.  Asterisks and arrows are defined as in Figure \ref{ew}.  We find strong correlations between both C II] and Mg II and $L_{acc}$.  Fits to the data are given in Equations \ref{lacccii} and \ref{laccmgii}.}
\label{cii}
\end{figure}

Included in our NUV spectra are two emission lines which have been shown to correlate with accretion properties.  \citet{calvet04} analyzed NUV spectra for a sample of intermediate mass T Tauri stars and found that luminosities of C II] $\lambda$2325 {\AA} and Mg II $\lambda$2800 {\AA} increased with $L_{acc}$.  We expected a similar trend due to simultaneous observations of the lines and accretion excess and Figure \ref{cii} shows a clear correlation.  Mg II is also a tracer of chromospheric activity, and is detected in each of the WTTS templates, but the chromospheric contribution to the luminosity is likely saturated at the age of our sample \citep{cardini07} and therefore not contributing to the spread in luminosity.  In the CTTS, Mg II can have strong absorption features due to the presence of outflows, which was seen in a sample of NUV spectra of T Tauri stars obtained with the International Ultraviolet Explorer \citep{ardila02}.  C II] is not observed in WTTS; however, it is easily identified even in low $\mdot$ objects and correlates with $L_{acc}$.  Therefore, C II] appears to be a very sensitive tracer of accretion or accretion related outflows \citep{calvet04,gomez05}, observable even when the NUV continuum excess is not.  Relations between line fluxes and $L_{acc}$ are given by;
\be
\label{lacccii}
\rm{log(L_{acc})=1.1(\pm0.2)log(L_{C\; II]})+2.7(\pm 0.7)}
\en
\be
\label{laccmgii}
\rm{log(L_{acc})=1.1(\pm0.2)log(L_{Mg\; II})+2.0(\pm 0.5)}
\en

In conclusion, there are several secondary tracers of accretion which are easily accessible in optical spectra and the good agreement between the correlations found here and in the literature, using different techniques, show that these correlations are robust.  With well calibrated accretion indicators, it is not necessary to go to the UV to obtain accretion rate estimates.  These results will facilitate the study of accretion properties for large samples of sources using optical spectra, which are significantly easier to obtain than UV data, and may offer more opportunities to measure $\mdot$ simultaneously to data tracing other T Tauri phenomena, like outflows or circumstellar disk dynamics.

\section{Summary and Conclusions}
We used a large sample of T Tauri stars with nearly simultaneous spectral coverage in the FUV, NUV and optical to analyze the accretion properties of young stars.  The main results are summarized here.

\begin{itemize}

\item
The NUV fluxes of WTTS are $\sim$3$\times$ stronger than dwarf stars of the same spectral type, due to enhanced chromospheric activity.  This emission can introduce uncertainties in the estimate of the NUV excess produced by accretion if the correct templates are not used.  By using WTTS templates with good signal to noise in the UV, which first became available in the DAO sample, we are able to distinguish chromospheric excesses from accretion excess when calculating accretion rates.  

\item
Accretion shock models which explain the NUV excess cannot describe both large filling factors predicted by models of the magnetic field distribution and non-zero 1 $\mu$m veiling.  Here, we modeled the shock emission as arising in columns carrying a range of energy fluxes (or densities) and covering different fractions of the stellar surface, to simulate the highly complex geometry of the accretion region.  The shock emission is assumed to be a combination of high energy flux ($\curf$) columns which peak in the UV and low $\curf$ accretion columns which peak at redder, optical wavelengths.  By including a variety of accretion columns, optical excesses and veiling at 1$\mu$m can be explained.  Also, the large filling factors of low energy columns are in better agreement with analysis based on magnetic field geometries.

\item
Comparing our estimates of the accretion properties to those calculated in previous analyses \cite{valenti93} and \citet{gullbring98}, we found that our assumptions for $A_V$ and intrinsic variability in the optical fluxes accounted for the majority of the differences in our $\mdot$ or $L_{acc}$ estimates.  Two factors in our technique appear to offset each other; 1) the UV flux is decreased by using a WTTS template and 2) fitting 1 $\mu$m veiling estimates increases the red shock excess.

\item
Low $\curf$ accretion columns which can explain 1 $\mu$m veiling in high accretors, may be present even when no red excess is observed.  The intrinsic stellar emission, mainly the photosphere in the red and the active chromosphere in the blue, may hide this accretion component at low $\mdot$.  We measured the maximum emission of a cool column that can be hidden by the stellar emission and determined the filling factor of the cool column.  We found large surface coverage, even in sources without veiling at 1 $\mu$m.  In some cases the upper limit on the red excess could be equal to the excess flux in the NUV, doubling $\mdot$.

\item
We used simultaneous measurements of emission line luminosities and accretion luminosities to calibrate correlations between the two. We found clear trends between the luminosities of H$\beta$, Ca II K, Mg II and CII].  No correlation is found between EW(H$\alpha$) and  $\mdot$, even with quasi-simultaneous observations, however the luminosity of H$\alpha$ is correlated with the accretion properties.  C II] may be a sensitive tracer of accretion at low $\mdot$, as it is observed in all the CTTS in our sample, even those with no NUV accretion excess, yet absent from the spectra of WTTS.

\end{itemize}

\section{Acknowledgments}
We would like to thank Melissa McClure for valuable discussions regarding extinction estimates and veiling in the near IR and providing results prior to publication.  We would also like to thank Ted Bergin, Lee Hartmann, Jon Miller and Fred Adams for providing comments on an early version of this paper as part of L. Ingleby's thesis. This work was supported by NASA grants for Guest Observer program 11616 to the University of Michigan, Caltech and the University of Colorado.  Based on observations made with the NASA/ESA {\it Hubble Space Telescope}, obtained from the Space Telescope Science Institute data archive. STScI is operated by the Association of Universities for Research in Astronomy, Inc. under NASA contract NAS 5-26555.  C.E. was supported by a Sagan Exoplanet Fellowship from NASA and administered by the NASA Exoplanet Science Institute.  RDA acknowledges support from the Science \& Technology Facilities Council (STFC) through an Advanced Fellowship (ST/G00711X/1).  SGG acknowledges support from STFC via a Ernest Rutherford Fellowship (ST/J003255/1).


%

\clearpage
\begin{deluxetable}{lccccc}
\tablewidth{0pt}
\tablecaption{WTTS Sample and Properties
\label{tabwtts}}
\tablehead{
\colhead{Object} &\colhead{$A_V$}& \colhead{Luminosity}& \colhead{Radius}& \colhead{Mass}& \colhead{SpT} \\
\colhead{}&\colhead{mag}&\colhead{($\lsun$)}&\colhead{($\rsun$)} &\colhead{$\msun$} & \colhead{}}
\startdata
HBC 427& 0.0&0.8&1.9& 0.8& K7\\
LkCa 19& 0.0&1.7&1.6 &1.3& K0\\
RECX 1&0.0&1.0&1.8&0.9&K5\\
TWA 7 & 0.0& 0.5& 1.8& 0.5& M1\\
\enddata
\tablecomments{$A_V$ and SpT references: HBC 427 and LkCa 19 \citep{kenyon95}; RECX 1  \citep{luhman04b}; TWA 7 \citep{webb99}.}
\end{deluxetable}

\begin{deluxetable}{lccccccccc}
\tablewidth{0pt}
\tablecaption{CTTS Sample and Properties
\label{tabctts}}
\tablehead{
\colhead{Object} &\colhead{SpT}&\colhead{$A_V$}&\colhead{Luminosity}& \colhead{Radius}& \colhead{Mass}&  \colhead{Distance}& \colhead{$r_V$}&\colhead{$r_Y$} \\
\colhead{}&\colhead{}&\colhead{}&\colhead{($\lsun$)}&\colhead{($\rsun$)} &\colhead{($\msun$)}&\colhead{(pc)} & \colhead{}&\colhead{}}
\startdata
AA Tau& K7&1.9&1.0&2.1&0.8&140& 0.3& 0.2\\
BP Tau& K7&1.1&1.0 &2.1&0.8&140& 0.7& 0.3\\
CS Cha&K6&0.3&1.9&2.7&0.9&160&\emph{0.2}&\emph{0.0}\\
CV Cha&G9&1.5&3.1&2.0&1.5&160&\emph{1.1}&\emph{0.6}\\
DE Tau&M2&0.9 &0.8&2.4&0.4 &140& 0.6&0.2\\
DK Tau A&K7&1.3 &1.6&2.6& 0.7&140& 0.5&0.5\\
DM Tau&M1&0.7&0.2&1.1& 0.5&140& \emph{0.7}&\emph{0.2}\\
DN Tau&M0&0.9&1.5&2.8& 0.6&140& 0.1&0.0\\
DR Tau&K5&1.4&0.4&1.1&0.9&140&8.1&2.0\\
FM Tau&M0&0.7&0.1&0.7&0.6&140& $>$2&\emph{0.8}\\
GM Aur&K7&0.6&1.2&2.3&0.8& 140&0.2&0.0\\
HN Tau A&K5&1.1&0.7&1.5&1.1&140& 0.8&0.5\\
IP Tau&M0&1.7&0.7&1.9&0.6  &140&\emph{0.2}&\emph{0.1}\\
LkCa 15&K5&1.1&0.8&1.6&1.1&140&\emph{0.2}&\emph{0.0}\\
PDS 66&K1&0.2&0.9&1.3&1.1&86&\emph{0.2}&\emph{0.1}\\
RECX 11&K5&0.0&0.6&1.4&1.0&97&\emph{0.0}&\emph{0.0}\\
RECX 15&M3&0.0&0.1&0.9&0.3&97&\emph{0.8}&\emph{0.4}\\
RW Aur A&K3&0.5&0.5&1.1&0.9 &140& 2.0&0.9\\
TWA 3a&M3&0.0&0.4&1.8&0.3&50&\emph{0.0}&\emph{0.0}\\
TW Hya&K7&0.0&0.3&1.1&0.8&56&0.3&0.0\\
V836 Tau&K7&1.5&1.0&2.1&0.8&140& 0.0&0.0\\
\enddata
\tablecomments{$A_V$ and SpT references: AA Tau, BP Tau, DE Tau, DN Tau, DR Tau, FM Tau, GM Aur, HN Tau A, IP Tau, LkCa 15, RW Aur A and V836 Tau  \citep{furlan11}; DK Tau A, DM Tau, CS Cha and CV Cha \citep{furlan09}; RECX 11 and RECX 15 \citep{luhman04b}; TWA 3a and TW Hya \citep{webb99}; PDS 66 \citep{mamajek02}.  Distance references: Taurus Molecular Cloud \citep{kenyon94}; Chamaeleon I \citep{whittet97}; $\eta$ Chamaeleon \citep{mamajek99}; TWA \citep{webb99}; PDS 66 \citep{mamajek02}.  $r_V$ and $r_Y$ values were taken from \citet{edwards06} and \citet{hartmann90} except for those sources identified in Section \ref{sectveil}.  $r_V$ and $r_Y$ from the literature are in regular font, whereas those determined from new observations or found assuming a relation between $r_V$ and $r_Y$  are in italics.   }
\end{deluxetable}

\begin{deluxetable}{lcccc}
\tablewidth{0pt}
\tablecaption{Log of Observations
\label{tabobs}}
\tablehead{
\colhead{Object} &\colhead{RA}& \colhead{DEC}&\colhead{Telescope/ Instrument}& \colhead{Date of Obs} \\
\colhead{}&\colhead{(J2000)}&\colhead{(J2000)}}
\startdata
AA Tau&04 34 55.42&+24 28 52.8&HST/ STIS G230L/G430L&2011-01-07\\
&&&CTIO/ SMARTS RC Spectrograph&2010-12-31\\
&&&CTIO/ SMARTS RC Spectrograph&2011-01-02\\
&&&CTIO/ SMARTS RC Spectrograph&2011-01-04\\
&&&CTIO/ SMARTS RC Spectrograph&2011-01-05\\
BP Tau&04 19 15.86&+29 06 27.2&HST/ STIS G230L&2002-01-12\\
CS Cha&11 02 25.20&-77 33 36.3&HST/ STIS G230L/G430L&2011-06-01\\
&&&CTIO/ SMARTS RC Spectrograph&2011-05-30\\
&&&CTIO/ SMARTS RC Spectrograph&2011-05-31\\
&&&CTIO/ SMARTS RC Spectrograph&2011-06-02\\
&&&Magellan/ MIKE& 2012-02-16\\
&&&VLT/ CRIRES&2010-12-23\\
CV Cha&11 12 27.65&-76 44 22.1&HST/STIS E230M/G430L&2011-04-13\\
&&&CTIO/ SMARTS RC Spectrograph&2011-04-08\\
&&&CTIO/ SMARTS RC Spectrograph&2011-04-13\\
&&&CTIO/ SMARTS RC Spectrograph&2011-04-15\\
&&&CTIO/ SMARTS RC Spectrograph&2011-04-17\\
&&&CTIO/ Bench-Mounted Echelle Spectrograph&2011-04-12\\
&&&Magellan/ MIKE& 2010-03-10\\
DE Tau&04 21 55.69&+27 55 06.1&HST/ STIS G230L/G430L&2010-08-20\\
&&&CTIO/ SMARTS RC Spectrograph&2010-08-15\\
&&&CTIO/ SMARTS RC Spectrograph&2010-08-18\\
&&&CTIO/ SMARTS RC Spectrograph&2010-08-20\\
&&&CTIO/ SMARTS RC Spectrograph&2010-08-21\\
DK Tau A&04 30 44.25&+26 01 24.5&HST/ STIS G230L/G430L&2010-02-04\\
DM Tau&04 33 48.74&+18 10 09.7&HST/ STIS G230L/G430L&2010-08-22\\
&&&CTIO/ SMARTS RC Spectrograph&2010-08-15\\
&&&CTIO/ SMARTS RC Spectrograph&2010-08-18\\
&&&CTIO/ SMARTS RC Spectrograph&2010-08-20\\
&&&CTIO/ SMARTS RC Spectrograph&2010-08-21\\
&&&Magellan/ MIKE& 2011-01-04\\
&&&VLT/ CRIRES&2011-08-11\\
DN Tau&04 35 27.44&+24 14 59.1&HST/ STIS G230L/G430L&2011-09-10\\
&&&CTIO/ SMARTS RC Spectrograph&2011-09-07\\
&&&CTIO/ SMARTS RC Spectrograph&2011-09-11\\
DR Tau&04 47 06.22&+16 58 42.6&HST/ STIS G230L/G430L&2010-02-15\\	
FM Tau&04 14 13.56&+28 12 48.8&HST/ STIS G230L/G430L&2011-09-21\\	
&&&CTIO/ SMARTS RC Spectrograph&2011-09-21\\
&&&CTIO/ SMARTS RC Spectrograph&2011-09-27\\	
&&&VLT/ CRIRES&2010-12-02\\
GM Aur&04 55 10.98&+30 21 59.1&HST/ STIS G230L/G430L&2010-08-19\\
HBC 427&04 56 02.02&+30 21 03.2&HST/ STIS G230L/G430L&2011-03-30\\
HN Tau A&04 33 39.37&+17 51 52.1&HST/ STIS G230L/G430L&2010-02-10\\
IP Tau&04 24 57.14&+27 11 56.4&HST/ STIS G230L/G430L&2011-03-21\\
&&&VLT/ CRIRES&2010-11-13\\
LkCa 15&04 39 17.73&+22 21 03.8&HST/ STIS G230L&2003-02-13\\
&&&Magellan/ MIKE& 2011-01-04\\
&&&VLT/ CRIRES&2010-02-03\\
&&&VLT/ CRIRES&2010-02-05\\
LkCa 19&04 55 36.97&+30 17 55.0&HST/ STIS G230L/G430L&2011-03-31\\
PDS 66&13 22 07.45&-69 38 12.6&HST/ STIS G230L/G430L&2011-05-23\\
&&&CTIO/ SMARTS RC Spectrograph&2011-05-16\\
&&&CTIO/ SMARTS RC Spectrograph&2011-05-20\\
&&&CTIO/ SMARTS CHIRON&2011-05-20\\
&&&CTIO/ Bench-Mounted Echelle Spectrograph&2011-05-22\\
RECX 1&08 36 56.12&-78 56 45.3&HST/ STIS G230L/G430L&2010-01-22\\
&&&VLT/ CRIRES&2011-05-30\\
RECX 11&08 47 01.28&-78 59 34.1&HST/ STIS G230L/G430L&2009-12-12\\
&&&CTIO/ SMARTS RC Spectrograph&2009-11-27\\
&&&CTIO/ SMARTS RC Spectrograph&2009-12-15\\
&&&CTIO/ SMARTS RC Spectrograph&2009-12-19\\
&&&Magellan/ MIKE& 2010-03-10\\
&&&VLT/ CRIRES&2011-05-30\\
RECX 15&08 43 18.43&-79 05 17.7&HST/ STIS G230L/G430L&2010-02-05\\	
&&&CTIO/ SMARTS RC Spectrograph&2010-02-01\\
&&&CTIO/ SMARTS RC Spectrograph&2010-02-04\\
&&&CTIO/ SMARTS RC Spectrograph&2010-02-06\\
&&&Magellan/ MIKE& 2010-03-11\\
RW Aur A&05 07 49.51&+30 24 04.8&HST/ STIS G230L/G430L&2011-03-25\\	
TWA 3a&11 10 27.80&-37 31 51.2&HST/ STIS G230L/G430L&2011-03-26\\
&&&CTIO/ SMARTS RC Spectrograph&2011-03-18\\
&&&CTIO/ SMARTS RC Spectrograph&2011-03-22\\
&&&CTIO/ SMARTS RC Spectrograph&2011-03-25\\
&&&CTIO/ SMARTS RC Spectrograph&2011-03-29\\
&&&Magellan/ MIKE& 2010-03-10\\
TWA 7&10 42 29.94&-33 40 16.7&HST/ STIS G230L/G430L&2011-05-05\\
TW Hya&11 01 51.95&-34 42 17.7&HST/ STIS G230&2002-05-10\\
V836 Tau&05 03 06.62&+25 23 19.6&HST/ STIS G230L/G430L&2011-02-05\\	
&&&CTIO/ SMARTS RC Spectrograph&2011-01-28\\
&&&CTIO/ SMARTS RC Spectrograph&2011-02-01\\
&&&CTIO/ SMARTS RC Spectrograph&2011-02-03\\
&&&CTIO/ SMARTS RC Spectrograph&2011-02-08\\	
\enddata
\end{deluxetable}

\begin{deluxetable}{lccccccccc}
\tablewidth{0pt}
\tablecaption{Literature $A_V$s for Taurus Sources
\label{tabav}}
\tablehead{
\colhead{Object} &\colhead{F09/11}& \colhead{KH95}&\colhead{G98}&\colhead{V93}& \colhead{F11}& \colhead{M13}}
\startdata
AA Tau&1.9&0.5&0.7&1.3&1.3&--\\
BP Tau& 1.1&0.5&0.5&0.9&1.8&0.6\\
DE Tau&0.9&0.6&0.6&1.7&--&0.9\\
DK Tau A&1.3&0.8&1.4&1.2&1.8&--\\
DM Tau&0.7&0.0&--&0.1&--&--\\
DN Tau&0.9&0.5&0.3&0.5&--&--\\
DR Tau&1.4&--&--&1.0&1.8&2.0\\
FM Tau&0.7&0.7&--&0.8&--&--\\
GM Aur&0.6&0.1&0.3&0.5&--&--\\
HN Tau A&1.1&0.5&0.7&1.0&3.1&--\\
IP Tau&1.7&0.2&0.3&--&--&--\\
LkCa 15&1.1&0.6&--&--&--&--\\
RW Aur A&0.5&--&--&1.2&--&--\\
V836 Tau&1.5&0.6&--&--&--&1.4\\
\enddata
\tablecomments{F09/11 \citep{furlan09,furlan11}, KH95 \citep{kenyon95}, G98 \citep{gullbring98}, V93 \citep{valenti93}, F11 \citep{fischer11}, M13 \citep{mcclure12}}
\end{deluxetable}

\begin{deluxetable}{lcccccccc}
\tablewidth{0pt}
\tablecaption{Results from Multi-Component Model Fits to UV and Optical Spectra
\label{tabacc}}
\tablehead{
\colhead{Object} &\colhead{$f(10^{10})$}&\colhead{$f(3\times10^{10})$}&\colhead{$f(10^{11})$}&\colhead{$f(3\times10^{11})$}&\colhead{$f(10^{12})$}&\colhead{$f_{total}$}&\colhead{$r_Y$}&\colhead{$\mdot\; (\msunyr)$}}
\startdata
AA Tau&0&0&0&0&0.002&0.002&0.0&1.5$\times10^{-8}$\\
BP Tau&0&0&0.02&0&0.002&0.022&0.2&2.9$\times10^{-8}$\\
CS Cha&0.008&0&0.003&0&0&0.011&0.0&5.3$\times10^{-9}$\\
CV Cha&0&0.4&0&0.02&0&0.42&0.6&5.9$\times10^{-8}$\\
DE Tau&0.1&0&0.0008&0.0007&0&0.1&0.3&2.8$\times10^{-8}$\\
DK Tau A&0.01&0&0.005&0.005&0&0.02&0.1&3.4$\times10^{-8}$\\
DM Tau&0.08&0&0&0.001&0.0006&0.082&0.3&2.9$\times10^{-9}$\\
DN Tau&0&0&0&0.002&0&0.002&0.0&1.0$\times10^{-8}$\\
DR Tau&0&0&0.3&0.07&0.006&0.37&1.6&5.2$\times10^{-8}$\\
FM Tau&0&0.07&0&0&0.001&0.071&0.8&1.2$\times10^{-9}$\\
GM Aur&0&0&0.001&0.003&0&0.004&0.0&9.6$\times10^{-9}$\\
HN Tau A&0&0&0&0.01&0.004&0.014&0.1&1.4$\times10^{-8}$\\
IP Tau &0&0&0&0&0.001&0.001&0.0&7.2$\times10^{-9}$\\
LkCa 15 &0&0&0.01&0.0007&0.0001&0.011&0.1&3.1$\times10^{-9}$\\
PDS 66&0&0&0&0&0.0001&0.0001&0.0&1.3$\times10^{-10}$\\
RECX 11&0&0&0.001&0&0&0.001&0.0&1.7$\times10^{-10}$\\
RECX 15&0.02&0.001&0&0.0007&8$\times10^{-5}$&0.022&0.3&8.0$\times10^{-10}$\\
RW Aur A&0&0.03&0.2&0&0&0.23&0.8&2.0$\times10^{-8}$\\
TWA 3a&0&0&0&0&8$\times10^{-6}$&8$\times10^{-6}$&0.0&9.8$\times10^{-11}$\\
TW Hya&0&0&0&0.0013&0.0013&0.0026&0.0&1.8$\times10^{-9}$\\
V836 Tau&0&0&0&0&0.0002&0.0002&0.0&1.1$\times10^{-9}$\\
\enddata
\tablecomments{The values in parentheses represents the energy flux ($\curf$) of each column in erg s$^{-1}$ cm$^{-3}$.}
\end{deluxetable}

\begin{deluxetable}{lcc}
\tablewidth{0pt}
\tablecaption{Accretion Rates from Literature for Taurus Objects
\label{tabacclit}}
\tablehead{
\colhead{Object} & \colhead{$\mdot$(G98)$^a$}& \colhead{$\mdot$(V93)$^b$} \\
\colhead{}&\colhead{($\msunyr$)}&\colhead{($\msunyr$)}}
\startdata
AA Tau&3.3$\times10^{-9}$&7.1$\times10^{-9}$\\
BP Tau&2.9$\times10^{-8}$&2.4$\times10^{-8}$\\
DE Tau&2.6$\times10^{-8}$&1.8$\times10^{-7}$\\
DK Tau A&3.8$\times10^{-8}$&6.1$\times10^{-9}$\\
DM Tau&--&2.9$\times10^{-9}$\\
DN Tau&3.5$\times10^{-9}$&1.3$\times10^{-9}$\\
FM Tau&--&7.7$\times10^{-9}$\\
GM Aur&9.6$\times10^{-9}$&7.4$\times10^{-9}$\\
HN Tau A&1.3$\times10^{-9}$&3.9$\times10^{-9}$\\
IP Tau &8.0$\times10^{-10}$&--\\
RW Aur A&--&3.3$\times10^{-7}$\\
\enddata
\tablecomments{\\
$^a$ \citet{gullbring98}\\
$^b$ \citet{valenti93}}
\end{deluxetable}

\begin{deluxetable}{lcccc}
\tablewidth{0pt}
\tablecaption{Maximum $f$ and $\mdot$ Assuming Hidden Accretion Emission
\label{tabacc2}}
\tablehead{
\colhead{Object}&\colhead{$\mdot^a\; (\msunyr)$}&\colhead{$f^a$}&\colhead{$\mdot^b\; (\msunyr)$}&\colhead{$f^b$}}
\startdata
AA Tau&1.5$\times10^{-8}$&0.002&1.6$\times10^{-8}$&0.08\\
BP Tau&2.9$\times10^{-8}$&0.022&3.3$\times10^{-8}$&0.2\\
DN Tau&1.0$\times10^{-8}$&0.002&1.7$\times10^{-8}$&0.06\\
DR Tau&5.2$\times10^{-8}$&0.37&5.6$\times10^{-8}$&0.8\\
GM Aur&9.6$\times10^{-9}$&0.004&1.3$\times10^{-8}$&0.04\\
HN Tau A&1.4$\times10^{-8}$&0.014&1.5$\times10^{-8}$&0.06\\
IP Tau &7.2$\times10^{-9}$&0.001&9.4$\times10^{-9}$&0.03\\
LkCa 15 &3.1$\times10^{-9}$&0.011&3.6$\times10^{-9}$&0.03\\
PDS 66&1.3$\times10^{-10}$&0.0001&3.8$\times10^{-10}$&0.02\\
RECX 11&1.7$\times10^{-10}$&0.001&2.6$\times10^{-10}$&0.006\\
TWA 3a&9.8$\times10^{-11}$&$8\times10^{-6}$&3.4$\times10^{-10}$&0.002\\
TW Hya&1.8$\times10^{-9}$&0.0026&2.3$\times10^{-9}$&0.05\\
V836 Tau&1.1$\times10^{-9}$&0.0002&2.6$\times10^{-9}$&0.02\\
\enddata
\tablecomments{\\
$^a$ Values which give the best $\chi_{red}^2$ fit of the model to the UV and optical data (see Table \ref{tabacc}).\\
$^b$ Values calculated when allowing for a low $\curf$ column which produces emission not detectable above the intrinsic stellar emission.\\}
\end{deluxetable}

\end{document}